# Tracking the Electron Transfer Cascade in European Robin Cryptochrome 4 Mutants

Daniel Timmer[a], Daniel C. Lünemann[a], Anitta R. Thomas[a], Anders Frederiksen[a], Jingjing Xu[b], Rabea Bartölke[b], Jessica Schmidt[b], Antonietta De Sio[a], Ilia A. Solov'yov[a,c], Henrik Mouritsen[b,c], Christoph Lienau[a,c].

[a] Institut für Physik, Carl von Ossietzky Universität, 26129 Oldenburg, Germany
[b] Institut für Biologie und Umweltwissenschaften, Carl von Ossietzky Universität, 26129 Oldenburg, Germany
[c] Research Centre for Neurosensory Science, Carl von Ossietzky Universität, 26111 Oldenburg, Germany

*Corresponding author(s): Christoph Lienau.

**Email:**  christoph.lienau@uni-oldenburg.de

**Abstract**

The primary step in the elusive ability of migratory birds to sense weak Earth-strength magnetic fields is supposedly the light-induced formation of a long-lived, magnetically sensitive radical pair inside a cryptochrome flavoprotein located in the retina of these birds. Blue light absorption by a flavin chromophore triggers a series of sequential electron transfer steps across a tetradic tryptophan chain towards the flavin acceptor. The recent ability to express cryptochrome 4 from the night-migratory European robin (*Erithacus rubecula*), *Er*Cry4, and to replace the tryptophan residues individually by a redox-inactive phenylalanine offers the prospect of exploring the role of each of the tryptophan residues in the electron transfer chain. Here, we compare ultrafast transient absorption spectroscopy of wild type *Er*Cry4 and four of its mutants having phenylalanine residues in different positions of the chain. In the mutants we observe that each of the first three tryptophan residues in the chain adds a distinct relaxation component (time constants 0.5, 30 and 150 ps) to the transient absorption data. The dynamics in the mutant with a terminal phenylalanine residue are very similar to those in wild type *Er*Cry4, excepted for a reduced concentration of long-lived radical pairs. The experimental results are evaluated and discussed in connection with Marcus-Hopfield theory, providing a complete microscopic insight into the sequential electron transfers across the tryptophan chain. Our results offer a path to studying spin transport and dynamical spin correlations in flavoprotein radical pairs.

**Introduction**

Cryptochromes are blue-light-sensitive flavoproteins that have a variety of signaling functions (1-4). They are essential, for example, for maintaining the circadian rhythm in animals (5, 6) and plants (7), and affect plant growth (8) and flowering (9). Moreover, there is increasing (10), albeit so far indirect, evidence that cryptochromes are involved in the elusive sensing of the Earth's magnetic field direction by migratory songbirds (11-14).

Cryptochrome proteins (Cry) from some species (10, 15, 16) internally bind a flavin adenine dinucleotide (FAD) chromophore non-covalently that bridges the protein surface through a chain of aromatic amino acid residues of approximately 25 Å. In plant Cry, the chain consists of three tryptophans (Trp) (12, 17), whereas animal and animal-like Crys typically have a chain of four Trps (10, 18). Blue light absorption via a $\pi \to \pi^*$ transition in the isoalloxazine moiety of the FAD cofactor in its fully oxidized ground state ($FAD_{ox}$) triggers a series of sequential electron transfers across the Trp chain and results in the formation of a pair of flavin/tryptophan radicals $[FAD^{\bullet-} TrpH^{\bullet+}]$ in a spin-correlated singlet state (10, 19, 20). The lifetime of this radical pair state ($^S$RP1), which is typically in the µs range (10, 11, 21), is expected to be long enough for a weak magnetic field to affect the quantum yield for intersystem crossing to its triplet state ($^T$RP1) (12, 22-24). Intersystem crossing is mainly induced by hyperfine couplings and thus depends on the alignment of nuclear spins (12, 18, 23). RP1 is expected to be sensitive to the direction of the external magnetic field and, therefore, is a likely candidate for the magnetically sensitive radical pair in birds (10). RP1 deprotonates on the ms time scale, thereby forming a magnetically insensitive, secondary radical pair, $[FAD^{\bullet-} Trp^{\bullet}]$ (21), followed by protonation of the flavin radical and reduction of the tryptophan radical. This leads to the formation of a signaling state with a magnetic field-sensitive yield that returns to the fully oxidized ground state on a much longer time scale.

Consequently, the ultrafast light-induced electron transfer across the Trp chain forms the primary step in the photochemical cycle of Crys. The ultrafast dynamics of flavins and flavoproteins have been the subject of numerous experimental studies during the past two decades (19, 25-33). In particular, for FAD in aqueous solution, an intramolecular electron transfer from the adenine to the isoalloxazine moiety occurs in stacked conformation with a time constant of 5 ps, followed by charge recombination on a faster time scale (30). In open conformation, a much longer, ns lifetime has been measured (30). For protein-bound FAD, a fast electron transfer from a nearby tryptophan has been observed that forms a $[FAD^{\bullet-} TrpH^{\bullet+}]$ radical pair within ~1 ps in animal type I Cry (19, 28, 34) and with sub-picosecond time constants in plant Cry and photolyases (27, 35-38). Dynamics on longer time scales are mostly interpreted in terms of the sequential electron transfer steps along the Trp chain. It remains difficult, however, to unambiguously assign rate constants to individual steps of the charge migration since competing nonequilibrium dynamics such as vibrational cooling (28) may affect the transient absorption spectra.

Very recently, it became possible to recombinantly express and purify (39) wild type cryptochrome 4 from the night-migratory European robin (10) (*Er*Cry4) with internally bound FAD chromophores with high yield (>97%) (10). Transient absorption spectra of wild type *Er*Cry4 and four mutants, in which one of the Trp involved in Cry activation has been site-selectively replaced by a redox-inactive phenylalanine (Phe) have been studied with sub-ns time resolution (10). The study gave evidence for the creation of light-induced radicals with lifetimes exceeding 100 ns in the wild type protein and in the mutant in which the terminal Trp has been substituted by the Phe residue. The experiments also showed sizable changes in the transient absorption by up to 15% for magnetic field strengths of ~10 mT. Due to time resolution limitations, the sequential charge transfer rates could not be resolved experimentally but were estimated computationally based on molecular dynamics simulations and empirical Moser-Dutton theory (10, 40-42). Since those rates are critical for understanding the magnetic sensitivity of *Er*Cry4, establishing them from independent ultrafast spectroscopic experiments is urgently needed. Such experiments would open new frontiers for probing intraprotein charge and spin transfer dynamics (43) and important horizons for experiment/theory comparison (10, 18, 44, 45).

In the present study, we provide a comparative ultrafast optical study of the electron transfer cascade along a tetradic amino acid chain in wild type *Er*Cry4 and in its four mutants, where one of the four tryptophans in the chain has been site-selectively replaced with a phenylalanine residue. This approach allows us to isolate the dynamics and yields of the first three sequential electron transfers across the chain in *Er*Cry4 and provides a benchmark for modeling radical pair formation in cryptochromes. Our study thus forms a basis for the investigation of spin transport and correlations in magnetically sensitive flavin-tryptophan radical pairs.

**Results and Discussion**

Spectroscopic studies were performed on wild type *Er*Cry4 and four mutants expressed and purified as described earlier (10). The atomistic structure of the wild type *Er*Cry4 protein has not been obtained experimentally, as it has not been crystallized yet. Gene sequencing (46) and computational modeling (10, 46) suggest that it is overall similar to that of pigeon (*Columba livia*, *Cl*) Cry4, also with four Trps involved in the activation, namely Trp$_A$ (W395), Trp$_B$ (W372), Trp$_C$ (W318), and Trp$_D$ (W369) (16). The structure of the FAD chromophore and the Trp tetrad, as obtained from such simulations, is schematically depicted in Fig. 1e. Site-specific mutagenesis (10) has been used to express four mutants, $W_XF$ (X=A, B, C, D), in which one of the four Trp (W) has been selectively replaced with phenylalanine (F) to block the electron transfer at different positions in the *Er*Cry4 structure, as illustrated in Figs. 1a-d.

The linear absorption spectra of wild type *Er*Cry4 and its four mutants are found to be very similar (Fig. 1f), featuring a main peak around 450 nm with well-resolved vibronic substructure. Recent hybrid quantum/classical modeling (47), taking vibronic couplings to nuclear vibrations explicitly into account, successfully described the substructure of these spectra and assigned the peak to a $\pi \rightarrow \pi^*$ transition from the $\pi_2$ to $\pi_3$ orbitals localized on the isoalloxazine moiety of the FAD cofactor inside Cry (48). The vibronic substructure is a distinct sign of chromophore binding to the protein and vanishes for FAD in solution. The structured peak around 370 nm arises from the $\pi_1 \rightarrow \pi_3$ transition of FAD (47).

All studied mutant samples show weak photoluminescence (PL) with an unstructured emission around 550 nm (see Supplementary Information Fig. S1). The corresponding excitation spectra are similar and independent of the emission energy (Fig. S1). In contrast to their linear absorption spectra, the excitation spectra recorded from all protein samples show an absorption peak with unresolved vibronic substructure around 450 nm and are similar to those seen for molecular FAD in buffer solution (Fig. S1). A comparison of the PL emission intensities of the mutant samples with that of molecular FAD in buffer solution indicates that, in all protein samples, >97% of all FAD molecules are bound and that the weak residual PL seen in Fig. S1 stems from a small amount of residual, unbound FAD. This observation is confirmed by time-resolved PL studies (Fig. 1g) showing for all mutants, except for $W_AF$, a 4-ns mono-exponential decay matching that of molecular FAD in buffer. The $W_AF$ mutant reveals an additional fast, resolution-limited emission component (Fig. 1g, inset). In this mutant, Trp$_A$, which is close to the FAD chromophore, is replaced with Phe, making a rapid electron transfer from Trp$_A$ to FAD impossible. Therefore, the optically excited $FAD_{ox}^*$ in $W_AF$ may be longer lived and its emission is not completely quenched by the presence of a nearby Trp$_A$. In all other mutants, instead, the Trp$_A$ electron donor is present, resulting in efficient PL quenching and thus no emission from $FAD_{ox}^*$. This observation already points to an efficient electron transfer from Trp$_A$ to FAD.

Differential transmission $\Delta T/T$ spectra were recorded under comparable experimental conditions for wild type *Er*Cry4 and the four considered mutants. Linearly polarized pump pulses with 30-fs duration centered at 450 nm were used to resonantly excite the $\pi_2 \rightarrow \pi_3$ transition of the FAD chromophore inside *Er*Cry4. Pump-induced changes in the transmission of the sample were probed using a broadband white light continuum. $\Delta T/T$ spectra for magic angle polarization between pump and probe are shown in Fig. 2a-d for the four mutants for delay times $t_W$ between the pulses of up to 1.5 ns. All datasets are subjected to a global data analysis and can be readily explained through a multi-exponential decay model, including a minimal set of decay associated difference spectra (DADS) (49). Each DADS spectrum (Fig. S7) describes components of $\Delta T/T$ that decay or rise with an associated time constant. Since such spectra are sometimes somewhat difficult to interpret, we also have calculated evolution associated difference spectra (EADS(49), see methods). Here, $EADS_1$ represents the $\Delta T/T$ spectra before the onset of incoherent consecutive relaxation steps in the sample while $EADS_n, n > 1$ reflects, in simplified terms, the shape of the $\Delta T/T$ spectrum after the (n-1)th relaxation process has been completed. The resulting EADS spectra are presented in Fig. 2e-h.

We first consider the $W_BF$ mutant (Figs. 2b and 2f), where the conceptually simplest, sub-ps electron transfer dynamics from $Trp_A$ to FAD is expected. The distinct signature of this electron transfer is the vanishing of the stimulated emission (SE) band around 550 nm in Fig. 2b on a 0.5 ps scale. The $\Delta T/T$ spectra in Fig. 2b at early time delays, represented by the first EADS spectrum ($EADS_1$) in Fig. 2f, are recorded before this electron transfer sets in. These $\Delta T/T$ spectra thus show pump-induced changes in the spectrum of the oxidized species, $FAD_{ox}$, only. The spectra are therefore not yet affected by the formation of radicals in the sample. Indeed, $EADS_1$ (blue line in Fig. 2f) can well be understood on the basis of the known optical transitions of $FAD_{ox}$ (Fig. 3a). The pump excitation promotes electrons to vibrationally excited states of $\pi_3$ and this gives rise to a positive ($\Delta T > 0$) ground state bleaching (GSB) contribution to the differential transmission with a spectrum matching the linear absorption of $FAD_{ox}$ (blue line in Fig. 3a). The pronounced vibronic substructure around 450 nm is evident in $EADS_1$ and is a distinct marker for the concentration of optically excited $FAD_{ox}^*$ chromophores (28). Rapid vibrational relaxation within $\pi_3$, complete within ≤200 fs, gives rise to a strongly red-shifted stimulated emission ($\Delta T > 0$) band ($\pi_3 \rightarrow \pi_2$) around 550 nm (see red line in Fig. 3a and Fig. S8 in the Supplementary Information). GSB below 400 nm is covered by a pronounced excited state absorption (ESA, $\Delta T < 0$) around 360 nm (with slight substructure around 390 nm) from the $\pi_3 \rightarrow \pi_5$ transition (47) (red lines in Figs. 3a,c). In addition, the $\Delta T/T$ spectra suggest broadband ESA of $FAD_{ox}^*$, spanning the range from 500 nm to 700 nm and a spectrally almost flat ESA between 400 and 470 nm that is more difficult to assign since ESA of $FAD_{ox}^*$ has not yet been modeled with high accuracy (red lines in Fig. 3a,c). Similar experimental ESA spectra have also been recorded by Kutta et al. but could not yet be well reproduced by quantum chemical calculations (28).

Importantly, the overall shape of $EADS_1$ in Fig. 2e-h is very similar in all investigated *Er*Cry4 mutants and also the relative amplitude of the different contributions remains basically unchanged. Since mutation alters the Trp chain with little effect on the FAD chromophore, this provides further support for the assignment of $EADS_1$ to the $\Delta T/T$ spectrum of the FAD chromophore in its oxidized form. The most obvious dynamic feature of the $\Delta T/T$ map for the $W_BF$ mutant (Fig. 2b) is the complete vanishing of the $\pi_3 \rightarrow \pi_2$ SE band around 550 nm with a decay time of 0.5 ps, while the amplitude of the GSB bands remains unchanged. This indicates that the decay of the excited state $\pi_3$ population is not due to the refilling of the $FAD_{ox}$ ground state $\pi_2$ but reflects the formation of the negatively charged $FAD^{\bullet-}$ radical, most likely due to the transfer of an electron from $Trp_A$. The map in Fig. 2b also shows more subtle features. After the first electron transfer has been completed, the $\Delta T/T$ spectrum, now given by $EADS_2$ in Fig. 2f, has changed substantially. Not only the characteristic SE band around 550 nm has disappeared completely, but also the sharp peak in the $FAD_{ox}^*$ absorption around 360 nm and the broad ESA tail (red line in Fig. 2f) has vanished in $EADS_2$. While the GSB and ESA around 450 nm do not change when switching from $EADS_1$ to $EADS_2$, a new narrow ESA band around 370 nm with a side peak at 400 nm and a broad ESA band in the 500-700 nm range emerge in $EADS_2$. After subtracting the GSB contribution from $EADS_2$, a $\Delta T/T$ spectrum is obtained that quantitatively matches the sum of the absorption spectra of the $FAD^{\bullet-}$ radical anion and the $TrpH^{\bullet+}$ radical cation known from the literature (28, 45). The corresponding spectra deduced from $EADS_2$ are shown as solid lines in Fig. 3b. The signatures of the spectrum of the Trp radical cation (purple line in Fig. 3c) are the broad absorption centered around 560 nm and a tail of a UV absorption band at 335 nm that is at the edge of the probe window (28, 36, 50). Theoretical studies of the $FAD^{\bullet-}$ absorption spectrum show a strong transition from the singly occupied $\pi_3$ to the unoccupied $\pi_5$ orbital (47), which are seen experimentally in the strong absorption peak in Fig. 3b (circles) at 370 nm, red-shifted by ca. 10 nm with respect to the corresponding transition in $FAD_{ox}$. Based on this theoretical analysis (47), we assign a weak peak at 400 nm in $EADS_2$ to the transition from $\pi_1 \to \pi_3$ and a broad resonance between 400 and 470 nm (green line in Fig. 3b) to the $\pi_2 \to \pi_3$ transition. Since the shape of the latter resonance is virtually identical to that seen in the ESA from $FAD_{ox}$, it is natural to assume that also in $FAD_{ox}$ this ESA peak reflects a two-quantum excitation of $\pi_2 \to \pi_3$. The decomposed spectra are in convincing agreement with those reported for other types of Cry, suggesting that the disappearance of $FAD_{ox}^*$ leads to the formation of the $[FAD^{\bullet-}Trp_AH^{\bullet+}]$ radical pair. The associated decay time of $\tau_1 = 1/k_1 = 0.51$ ps thus appears to be the time constant for the first electron transfer in *Er*Cry4 activation, corresponding to the $[FAD_{ox}^*Trp_AH] \xrightarrow{k_1} [FAD^{\bullet-}Trp_AH^{\bullet+}]$ process. Since, within our measurement precision, the characteristic markers for $FAD_{ox}^*$, i.e., the SE emission band around 550 nm and the ESA peak around 360 nm, are completely absent in $EADS_2$, we conclude that the yield for this first electron transfer $\eta_1$ is close to unity.

Our data analysis reveals some dynamics associated with a second time constant of $\tau_2 = 1.8$ ps. These dynamics processes, however, have only a minor effect on the $\Delta T/T$ spectra. The $EADS_3$ (green line in Fig. 2f), monitoring $\Delta T/T$ after the completed relaxation process, shows overall the same spectral structure and amplitude as $EADS_2$ (red line in Fig. 2f). Close inspection reveals a slight narrowing of the $TrpH^{\bullet+}$ band around 560 nm and a minor reshaping of the $FAD^{\bullet-}$ peaks at 370 and 400 nm. Some authors have recently associated this spectral evolution with vibrational cooling of the optically formed radicals (28) while others have assigned it to a second electron transfer across the Trp chain (45). The residual $\Delta T/T$ signal decays almost entirely with a time constant of $\tau_r = 89$ ps. This decay monitors the decay of the optically created radical pairs and the refilling of ground state $FAD_{ox}$. It is thus associated with the geminate recombination of the radical pairs. The residual, faint $EADS_4$ spectrum in Fig. 2f, with a shape matching that of $EADS_3$, yet with 20 times smaller amplitude, suggests that a small concentration of radical pairs is still possible for $t_W \gg \tau_r$. Importantly, the $\Delta T/T$ spectra of $W_BF$ show no clear signatures of other electron transfer processes except for $k_1$.

Ideally, one expects the first electron transfer in *Er*Cry4 activation to be efficiently blocked by substituting $Trp_A$ with Phe, as is done in the $W_AF$ mutant. Initially, the $\Delta T/T$ spectra in Fig. 2a ($EADS_1$, blue line in Fig. 2e) are very similar to those in the $W_BF$ mutant. The fast dynamics on the sub-ps scale, however, are no longer observed, which directly confirms the assignment of the first electron transfer in the dynamics of the $W_BF$ mutant. The fastest electron transfer dynamics in the case of the $W_AF$ mutant occurs with a time constant of 3.3 ps and results ($EADS_2$, red line in Fig. 2e) in a partial reduction of SE and ESA from $FAD_{ox}^*$. Since intermolecular electron transfer processes involving $Trp_A$ can not be the origin of this decay, the observed dynamics is most likely of intramolecular nature. Indeed, intramolecular photo-induced electron transfer from the adenine to isoalloxazine moiety of the FAD cofactor is known to occur in solution on a 5 ps time scale for the stacked FAD conformation (26, 30). Even though the exact time scale for the back electron transfer to adenine is somewhat debated for molecular FAD in solution (26, 30), it is generally thought to happen on a similar time scale as the forward transfer, forming an equilibrium between neutral and charge-separated species. The existence of such an equilibrium for the intramolecular electron transfer in the $W_AF$ mutant may be the reason why the recorded data show only a partial decay of $FAD_{ox}^*$ with a 3.3 ps time constant. More surprising is the $\Delta T/T$ dynamics with a larger 12 ps time constant. After its completion, the resulting $EADS_3$ (green line in Fig. 2a) matches that seen in the $W_BF$ mutant, indicating the formation of a $[FAD^{\bullet-}Trp_XH^{\bullet+}]$ radical pair. Molecular dynamics simulations of the protein structure (Fig. 5a) indicate that two additional Trp residues, Trp290 and Trp350, are present at an edge-to-edge distance of 4 Å from the adenine and 7.5 Å from the flavin moiety, respectively. These residues could then potentially be involved in the formation of a [FAD$^{\bullet-}$TrpH$^{\bullet+}$] radical pair. The $\Delta T/T$ changes in Fig. 2a disappear almost completely on a 100 ps time scale, the decay time associated with EADS$_3$. We can thus associate this 100 ps time constant with an effective geminate recombination of the formed intermolecular radical pair. Altogether, the experiments on the W$_A$F mutant prove that the first electron transfer in ErCry4 from Trp$_A$ is efficiently blocked by mutation. At the same time, our findings highlight the role of competing intra- and intermolecular electron transfer processes, involving electron donors outside the conserved tetradic Trp chain, for the radical pair formation in Cry.

The performed analysis puts us now in an excellent position to explore the electron transfer across the Trp chain in ErCry4 W$_C$F and W$_D$F mutants (Figs. 2c and 2d). In both cases, the $\Delta T/T$ spectra at early time delays (EADS$_1$ in Figs. 2g and 2h) are very similar to those discussed before for the W$_A$F and W$_B$F mutants, except for minor changes in peak amplitudes. In the case of the W$_C$F and W$_D$F samples, the initial dynamics points to a first electron transfer from Trp$_A$ to FAD with close to 100% efficiency and time constants of $\tau_1 = 0.49$ ps (W$_C$F) and 0.4 ps (W$_D$F), respectively. Also, the few-ps decay of EADS$_2$ from the W$_C$F mutant is seen, now with time constants of $\tau_2 = 4.1$ ps (W$_C$F) and 2.2 ps (W$_D$F). In contrast to the W$_B$F mutant case, the shape of the spectrum of the Trp radical cation changes slightly when going from EADS$_2$ to EADS$_3$, most dominantly around 500 nm and in the far red region. Since EADS$_2$ is not affected by an extension of the Trp chain, our data do not support the assignment of $\tau_2$ to an electron transfer process. Instead, our observation presents strong arguments in favor of vibrational cooling as suggested earlier (28). The spectra recorded for both the W$_C$F and the W$_D$F mutants show a third EADS$_3$ component with a spectral lineshape and a decay time ($\tau_3 = 58$ ps for W$_C$F and 26 ps for W$_D$F) that are very similar to those observed for the W$_B$F mutant. In stark contrast to the W$_B$F mutant, however, a fourth EADS$_4$ component now appears after the third relaxation step with an amplitude which is approximately 2/3 of that of EADS$_3$, and with an almost identical lineshape. This finding implies that, even after the third relaxation step, i.e., after about 100 ps, the excited ErCry4 in the sample still contains a sizeable concentration of radical pairs and that geminate recombination is not yet complete. This observation thus points to a secondary electron transfer step, $[\text{FAD}^{\bullet-}\text{Trp}_A\text{H}^{\bullet+}] \xrightarrow{k_2} [\text{FAD}^{\bullet-}\text{Trp}_B\text{H}^{\bullet+}]$, in which an electron is transferred from Trp$_B$ to Trp$_A$ with a rate constant $k_2 = 1/\tau_3$. The data in Figs. 2g and 2h indicate that this secondary electron transfer occurs with a yield of $\eta_2 \simeq 0.77$ for W$_C$F and 0.71 for W$_D$F. In the case of the W$_C$F mutant, the resulting radical pair concentration then recombines very slowly with $\tau_r \simeq 1.6$ ns. In the W$_D$F mutant, both the GSB and the radical pair absorption decay more quickly than for the other mutants. The analysis in Fig. 2h reveals the presence of a new decay channel with a decay time $\tau_4 = 144$ ps. This relaxation process has little effect on the shape of the $\Delta T/T$ spectra since the spectra from EADS$_3$ to EADS$_5$ differ only in amplitude. We assign the observed decay to the third electron transfer step, $[\text{FAD}^{\bullet-}\text{Trp}_B\text{H}^{\bullet+}] \xrightarrow{k_3} [\text{FAD}^{\bullet-}\text{Trp}_C\text{H}^{\bullet+}]$, in which an electron is transferred from Trp$_C$ to Trp$_B$ at a rate of $k_3 = 1/\tau_4$ and with a yield of $\eta_3 \simeq 0.52$. This yield is significantly lower than that of the second step. The population dynamics of the different neutral and charged radical states in mutant W$_D$F that is predicted by the analysis of the $\Delta T/T$ spectra is summarized in Fig. 3d. This analysis indicates that only a small fraction of about 20 – 25% of the initially generated radical pairs is still present after the first three electron transfers along the Trp chain (yellow line in Fig. 3d). These radical pairs then undergo geminate recombination on a time scale of $\tau_r = 2.0$ ns (EADS$_5$). Comparative studies of several samples of the same mutant indicate that the number of decay components needed to explain the experimental data do not depend on the specific studied sample. Also, the deduced decay times and yields are quite similar.

We now compare the results in Fig. 2 to a transient absorption study of wild type ErCry4 under similar experimental conditions. Qualitatively, the $\Delta T/T$ spectra in Fig. 4a look almost identical to those recorded for the W$_D$F mutant. Spectra at selected $t_W$ are shown in Fig. 4b. The dynamics at selected probe wavelengths are compared to the results from a global fit analysis presented in Fig. 4c. As in the case of the W$_D$F mutant, this analysis reveals five distinct EADS spectra with features that closely resemble those of the $W_DF$ mutant. Also, the decay times associated to those five components are almost the same as for the $W_DF$ mutant. This strongly indicates that the sequential electron transfer model developed from the mutant spectra can be used to explain the wild type data. As such, we assign the time constant $\tau_2 = 3$ ps to vibrational cooling and identify the decay time of $EADS_5$ with geminate recombination dynamics of the radical pair with $\tau_r = 2.2$ ns.

Our experimental results provide clear evidence for three sequential electron transfers in wild type *Er*Cry4, with decay constants of $k_1 = 1/0.39$ ps$^{-1}$, $k_2 = 1/30$ ps$^{-1}$ and $k_3 = 1/141$ ps$^{-1}$. The associated yields of the individual steps are $\eta_1 = 1.0$, $\eta_2 = 0.78$ and $\eta_3 = 0.77$. In these measurements, the yield of the third step is somewhat higher than in the case of the $W_DF$ mutant. This points to a radical pair concentration in the wild type protein being more than half of the initially optically generated concentration. The electron transfer rates and yields deduced for the wild type protein and all mutants are summarized in Table 1.

The measured electron transfer rate constants can now be compared to the values from theoretical models. Here, we have estimated the electron transfer rate constants within Marcus-Hopfield (MH) theory (51). The MH theory can be summarized in the following equation describing the electron transfer rate constant between a donor and an acceptor site within the protein:

$$k_{et}^{MH} = \frac{2\pi}{\hbar} V_0^2 e^{-\beta(R-R_0)} \frac{1}{\sqrt{2\pi\lambda\hbar\omega\coth\left(\frac{\hbar\omega}{2k_BT}\right)}} e^{-\frac{(\Delta G+\lambda)^2}{2\lambda\hbar\omega\coth\left(\frac{\hbar\omega}{2k_BT}\right)}}. \quad (1)$$

Here, $\hbar$ is the reduced Planck constant, $k_B$ is the Boltzmann constant and *T* is the temperature of the system. In Eq. (1), $V_0$ denotes the electronic coupling at a distance $R_0$ for which the electron coupling is known, and $\beta$ is the coefficient describing the decay of coupling for electrons inside a protein. *R* is the edge-to-edge distance between the donor and acceptor sites. The parameter $\lambda$ measures the reorganization energy required to distort the nuclear configuration of the protein into the product state without an electron transfer. The value $\Delta G$ is the free energy difference between the reactant and product states, and $\hbar\omega$ is the energy of the vibrational mode that couples the product and reactant states. The hyperbolic cotangent function, coth, becomes $2k_BT/(\hbar\omega)$ in the limit of $2k_BT \gg \hbar\omega$, transforming Eq. (1) to the well-known Marcus equation (52) for an electron transfer.

Here we have assumed a vibrational energy of $\hbar\omega = 50$ meV, following from the Moser-Dutton theory (40-42). The temperature has been set to 274 K according to the experimental conditions, and $R_0 = 3.6$ Å represents the van der Waals contact distance for the studied redox centers. The values for *R*, $\Delta G$ and $\lambda$ for *Er*Cry4 have been recently computed (10) using a combination of classical MD and quantum chemistry methods. Specifically, for ET$_2$ $R = 3.7$ Å, $\Delta G = -0.2521$ eV and $\lambda = 0.7675$ eV were used; for ET$_3$ $R = 3.7$ Å, $\Delta G = -0.3532$ eV and $\lambda = 1.336$ eV, whereas $R = 3.6$ Å, $\Delta G = -0.0311$ eV and $\lambda = 0.9887$ eV were assumed for ET$_4$.

The estimated electron transfer times, calculated from Eq. (1), are summarized in Fig. 5. They depend strongly on the electronic coupling ($V_0$) and less on $\beta$. In many cases earlier in the literature, $\beta$ was assumed equal to 1.4 Å$^{-1}$ (10, 40-42), to mimic a realistic protein environment. This choice is however more phenomenological and could in principle change somewhat for different proteins (53). The comparison in Fig. 5 suggests that a reasonable match of the experimental results with predictions of the theory within the MH approximation could be achieved for coupling strengths $V_0 \sim 2-4$ meV for ET$_2$, and $V_0 \sim 10-13$ meV for ET$_3$. All these values seem to be in the range expected for proteins (53). In general, the MH formulation may describe the electron transfer rates qualitatively well provided that the critical input parameters, in particular the coupling strength and the potential energy surfaces of the relevant donor and acceptor states, are reasonably well known.

Here, our experimental results provide a strong benchmark for advanced modelling of potential energy surfaces in Cry proteins. When using the parameters reported above for modelling the rate of the fourth electron transfer step in the chain, we would expect transfer times in the range of several hundreds of ps up to >1 ns. This is substantially longer than the third transfer time and at the edge of the time window that is probed in the present experiments.

A comparison of the electron transfer rate estimates in the present study with the ones reported earlier (10), also permits to conclude that the Moser-Dutton theory, used previously, is within an order-of-magnitude agreement with the results of the present measurements (30 ps for $ET_2$ and 140 ps for $ET_3$). The measured transfer times are somewhat slower than recently predicted (10). The electron transfer times presented in Fig. 5 illustrate that a better agreement could be achieved through the MH theory by choosing a more appropriate electronic coupling coefficient, rather than the generic one used in the Moser-Dutton approximation.

**Summary and conclusion**

The ability to express high purity cryptochrome 4 proteins from night-migratory European robins and to replace the tryptophan residues individually by redox-inactive phenylalanine has allowed us to provide new microscopic insight into the light-induced electron transfer dynamics across the tetradic tryptophan chain that connects the flavin chromophore to the surface of this protein. Using ultrafast optical spectroscopy, in connection with a Marcus-Hopfield model for the electron transfer inside the protein, we could unravel several of the primary steps in the photocycle of this photoreceptor which is suggested to play a decisive role in the magnetic compass sensing of migratory songbirds (10, 11, 14).

When replacing the tryptophan amino acid that is located most closely to the FAD chromophore ($Trp_A$) with a phenylalanine blocker, intermolecular electron transfer along the Trp tetrad is efficiently suppressed and slower intra- and intermolecular transfer paths are revealed. Molecular dynamics simulations suggest that the intermolecular steps likely involve other Trps, outside the tetrad. In all other mutants and in the wild type protein, $Trp_A$ serves as an efficient donor, transferring an electron to FAD within 0.5 ps after photoexcitation and with a yield of unity. The experiments reveal vibrational cooling dynamics within the manifold of radical pair states on a few-ps timescale which are insensitive to mutation. They provide evidence for a cascade of sequential electron transfers along the chain in which the second ($Trp_B \rightarrow Trp_A$) and third ($Trp_C \rightarrow Trp_B$) steps occur with transfer times of 30 ps and 140 ps, respectively, and can be selectively blocked by mutation. Assuming the validity of the Marcus-Hopfield model (Eq. (1)), we would expect that the electron transfer times decrease only slight, by approximately 50%, when increasing the temperature from 1°C, chosen in the present work to 42°C, the body temperature of European robins. As such, we expect that the main conclusions of this work remain valid under physiological conditions. An experimental study of the temperature dependence of the electron transfer dynamics is ongoing.

In the wild type protein, the yield is found to be quite high (> 75 %) while in the mutation $W_DF$, in which the terminal $Trp_D$ is replaced, the third electron transfer step has a lower yield, (~ 50%). Our data, probing the dynamics within the first 1.5 ns after photoexcitation, do not provide clear indication for a fourth transfer step from $Trp_D$ to $Trp_C$, either due to the lack of a good optical sensor for these distant tryptophans in the present measurements or because this step is either too slow or too fast to be resolved here. A comparison of the transient absorption data from the wild type protein and the mutant $W_DF$ suggests that the presence of the fourth tryptophan in the chain has very little effect on the dynamics within the first 1.5 ns, except for apparently increasing the concentration of radical pairs at the end of the measurement window. Such a change in concentration could be explained, e.g., by a transfer time for the fourth electron transfer from $Trp_D \rightarrow Trp_C$ that is faster than that for the third step and is forming an equilibrium between the associated radical pairs (10, 18). The present data indeed point to the formation of such an equilibrium and suggest more refined studies of the functional role of the terminal Trp.

While the time constant that we have deduced for the first and second electron transfers agree well with those recently reported by Kutta et al. for cryptochrome from fruit flies (28), the electron transfer times measured here are somewhat slower than recent predictions (10) on the basis of molecular dynamics simulations and Moser-Dutton theory. A first comparison of our data to predictions from Marcus-Hopfield theory indicates that they can be rationalized by taking slightly weaker electronic couplings than previously assumed for this protein (10). These simulations suggest that the fourth electron transfer step occurs with ~ 1 ns, beyond the window of the present experiments. Since the theoretical results depend quite sensitively on the atomistic and electronic structure of the protein, the parameters that enter the model are challenging to predict with high precision, even when using state-of-the art simulations (53). The experimental results presented here for the first three electron transfers in the Trp chain therefore are an important benchmark for future theoretical work.

Our results provide new microscopic insight into the inner workings of this sequential electron transfer chain. They demonstrate how charge flow and thus radical pair formation across the tryptophan tetrad in cryptochrome proteins can be sensitively controlled by site-selective mutation of the amino acid structure of the chain. This opens up a new and exciting avenue for probing not only the flow of charge but also the transport of spins (43) along such chains by all-optical means. This approach may be helpful for monitoring dynamical spin correlations in radical pairs in biomolecules or organic photovoltaic materials.

**Materials and Methods**

**Sample preparation.** Wild type *Er*Cry4 (GenBank: KX890129.1) and its four mutants $W_XF$ (X=A-D) were cloned, expressed and purified according to the protocol described by Xu et al. (10). Briefly, the tryptophan mutants were generated by replacing the DNA codon for tryptophan (TGG) at amino acid position 395 ($W_AF$), 372 ($W_BF$), 318 ($W_CF$), or 369 ($W_DF$) in the *Er*Cry4 gene by a phenylalanine codon (TTT) in a PCR using the Q5 site-directed mutagenesis kit (New England Biolabs). Plasmids were confirmed by Sanger sequencing (LGC Genomics). Proteins were expressed in BL21(DE3) *E. coli* cells in the dark and purified by Ni-NTA agarose columns, followed by anion exchange chromatography. Purified protein samples were concentrated to 5-6 mg/ml in an aqueous buffer solution (20mM Tris, 250mM NaCl and 20% glycerol) along with 10 mM of the reducing agent 2-mercaptoethanol (BME) to avoid dimerization of the protein. Samples were snap frozen in liquid nitrogen and stored at -80°C for 3 – 5 days until the measurements. Since the photocycle starts from the fully oxidized state $FAD_{ox}$, the reducing agent was removed and the sample was fully oxidized prior to the optical measurements. For this, the protein sample was washed with BME-free buffer solution in a Millipore centrifugation filter (Amicon Ultra, 30 kDa) using a temperature-controlled microcentrifuge (4°C, 14000 rpm). This step also removed free FAD from the sample. Addition of 1.5 mM potassium ferricyanide (PFC), followed by centrifugation for one hour to remove aggregated proteins, was repeated until the sample was fully oxidized as confirmed by absorbance measurements. A remaining concentration of ~1.7-2.6 mM PFC prevented photoreduction during the optical experiments. The PFC-corrected absorbance was used to determine the final concentration of the samples to be ~120-220 µM using the molar extinction of $FAD_{ox}$ (30). During the preparation the samples were always kept at ~4°C and measurements were performed at 1°C. Transient absorption measurements on a buffer solution containing 1 mM PFC showed only very weak nonlinear signal with a few-ps lifetime when pumped at 450 nm (see Fig. S4 in the Supplementary Information) and no detectable photoluminescence was observed.

**Photoluminescence (PL) experiments.** PL measurements were performed using broadband (400-680 nm) 1.25 pJ pulses from a 40 MHz fiber laser source (SC400, Fianium). Emitted light is collected with a microscope objective with numerical aperture of 0.3 and detected using a single-photon avalanche photodiode and a time-correlated single photon counting unit (PicoHarp 300, PicoQuant). The instrument response function (IRF) of the detection system at 550 nm is ~60 ps. To obtain correlated excitation and emission spectra, a Fourier transform approach using two phase-stable common-path interferometers (Translating Wedge-based Identical pulse eNcoding System, TWINS (54)) is used (55), yielding a spectral resolution of ~6 nm in excitation and emission wavelength. A more detailed description of the setup and data analysis can be found in the Supplementary Information.

**Transient absorption experiments.** The measurements were performed using ~30 fs pump pulses at 450 nm on a 15 µl volume of the sample in a quartz microcuvette (Hellma). The pulses are generated in an optical parametric amplifier (Topas, Light Conversion) pumped by regeneratively amplified 25-fs pulses at 800 nm (Legend Elite, Coherent) at a repetition rate of 10 kHz. Differential transmission spectra $\Delta T(\lambda, t_W)/T = \left(T_{\text{on}}(\lambda, t_W) - T_{\text{off}}(\lambda)\right)/T_{\text{off}}(\lambda)$ were recorded as a function of probe wavelength $\lambda$ and time delay $t_W$ using a broad supercontinuum probe, generated in a CaF$_2$ crystal. Here, $T_{\text{on/off}}$ denotes the probe transmission in the presence/absence of the pump, respectively. The spectra have been recorded with parallel and crossed polarizations of the pump and probe beams. Scattering corrections to the $\Delta T$ spectra were made as described in the SI. The experiments were performed with 20 nJ pump pulses focused to a spot size of ~50 µm and it was ensured that the $\Delta T/T$ signals are well in the linear regime (see Fig. S2). No change in signal due to sample degradation could be observed during the measurements (see Fig. S3).

**Data analysis.** Transient absorption scans recorded with parallel and crossed pump and probe polarizations were averaged and isotropic, magic angle spectra were calculated. The few-ps chirp of the probe continuum was corrected by extracting the wavelength-dependent time delay zero ($t_W(\lambda) = 0$) from the cross-phase modulation artifact of a transient absorption measurement of plain buffer solution. The first 200 fs of the corrected dynamics were discarded due to residual coherent signal contributions from the solvent. The datasets were subjected to a global analysis using a multi-exponential decay model (56). This decomposes the data into a set of *n* decay associated difference spectra (DADS$_i$) with corresponding decay times $\tau_i$, $\Delta T/T(\lambda, t_W) = \sum_{i=1}^{n} DADS_i(\lambda)\, e^{-t_W/\tau_i}$. The lowest number of decays necessary to simultaneously reproduce the data at all wavelengths was taken. For more details see the Supplementary Information. The DADS spectra were then used to obtain evolution associated difference spectra $EADS_k(\lambda) = \sum_{i=k}^{n} DADS_i(\lambda)$.

**Molecular dynamics simulations.** Using the amino acid sequence from an earlier study (10, 46), a computational structure for the *Er*Cry4 was constructed utilizing the SWISS-model workspace and pigeon Cry4 (PD ID: 6PU0) as a template. The homology model for the dark state *Er*Cry4 structure was solvated, neutralised and energy-minimized for 10000 conjugate-gradient steps. The structure was dynamically equilibrated for 5 ns followed by 200 ns production simulation. Four radical state simulations were created succeeding the dark state simulation, one for each of the radical pair states of the tryptophan residues associated with the electron transfers. All simulations were carried out using the CHARMM36 force fields for proteins with CMAP corrections, water and ions as well as special parameters created for FAD$_{\text{ox}}$, FAD$^{\bullet-}$, and TrpH$^{\bullet+}$, the latter has been successfully employed in earlier studies (57, 58).

All simulation runs utilized an integration timestep of 2 fs and had a temperature control using a Langevin thermostat with a damping coefficient of 5.0 ps$^{-1}$ applied for all heavier atoms in the system. The pressure was similarly held constant at 1 atm utilizing a Langevin Barostat. The ShakeH algorithm was used to keep hydrogen atoms at a fixed position and periodic boundary conditions were used with long-range Coulomb interactions being treated using the Particle Mesh Ewald summation method. The van der Waals interactions were calculated using a smooth cut-off distance of 12 Å with a switching distance of 10 Å. All MD analysis on the system were obtained utilizing VMD 1.9.3 and the VIKING web interface (59).

**Acknowledgments**

Financial support by the Deutsche Forschungsgemeinschaft (project number 395940726-SFB1372: Magnetoreception and Navigation in Vertebrates: From biophysics to brain and behaviour, GRK1885 "Molecular basis of sensory biology", INST 184/163-1 FUGG, Li 580/16-1 and DE 3578/3-1), by the European Research Council (under the European Union's Horizon 2020 research and innovation program, grant agreement no. 810002 (Synergy Grant: "QuantumBirds")) and by the Volkswagen Foundation (SMART; Lichtenberg professorship) is gratefully acknowledged. Computational resources for the simulations were provided by the CARL Cluster at the Carl-von-Ossietzky University, Oldenburg, supported by the DFG and the Ministry for Science and Culture of Lower Saxony. This work was also supported by the North-German Supercomputing Alliance (HLRN). We thank P.J. Hore and K.B. Henbest for fruitful discussions.

**References**

1. I. Chaves *et al.*, The Cryptochromes: Blue Light Photoreceptors in Plants and Animals. *Annu Rev Plant Biol* **62**, 335-364 (2011).
2. E. A. Griffin, D. Staknis, C. J. Weitz, Light-Independent Role of CRY1 and CRY2 in the Mammalian Circadian Clock. *Science* **286**, 768-771 (1999).
3. D. S. Hsu *et al.*, Putative Human Blue-Light Photoreceptors hCRY1 and hCRY2 Are Flavoproteins. *Biochemistry* **35**, 13871-13877 (1996).
4. H. Zhu *et al.*, Cryptochromes Define a Novel Circadian Clock Mechanism in Monarch Butterflies That May Underlie Sun Compass Navigation. *PLOS Biology* **6**, e4 (2008).
5. R. Stanewsky *et al.*, The cry(b) mutation identifies cryptochrome as a circadian photoreceptor in Drosophila. *Cell* **95**, 681-692 (1998).
6. G. T. J. van der Horst *et al.*, Mammalian Cry1 and Cry2 are essential for maintenance of circadian rhythms. *Nature* **398**, 627-630 (1999).
7. D. E. Somers, P. F. Devlin, S. A. Kay, Phytochromes and cryptochromes in the entrainment of the Arabidopsis circadian clock. *Science* **282**, 1488-1490 (1998).
8. C. T. Lin *et al.*, Enhancement of blue-light sensitivity of Arabidopsis seedlings by a blue light receptor cryptochrome 2. *P Natl Acad Sci USA* **95**, 2686-2690 (1998).
9. H. W. Guo, W. Y. Yang, T. C. Mockler, C. T. Lin, Regulations of flowering time by Arabidopsis photoreceptors. *Science* **279**, 1360-1363 (1998).
10. J. J. Xu *et al.*, Magnetic sensitivity of cryptochrome 4 from a migratory songbird. *Nature* **594**, 535-+ (2021).
11. P. J. Hore, H. Mouritsen, The Radical-Pair Mechanism of Magnetoreception. *Annu Rev Biophys* **45**, 299-344 (2016).
12. I. A. Solov'yov, D. E. Chandler, K. Schulten, Magnetic Field Effects in Arabidopsis thaliana Cryptochrome-1. *Biophys J* **92**, 2711-2726 (2007).
13. A. Pinzon-Rodriguez, S. Bensch, R. Muheim, Expression patterns of cryptochrome genes in avian retina suggest involvement of Cry4 in light-dependent magnetoreception. *Journal of The Royal Society Interface* **15**, 20180058 (2018).
14. H. Mouritsen, Long-distance navigation and magnetoreception in migratory animals. *Nature* **558**, 50-59 (2018).
15. R. J. Kutta, N. Archipowa, L. O. Johannissen, A. R. Jones, N. S. Scrutton, Vertebrate Cryptochromes are Vestigial Flavoproteins. *Sci Rep-Uk* **7**, 44906 (2017).
16. B. D. Zoltowski *et al.*, Chemical and structural analysis of a photoactive vertebrate cryptochrome from pigeon. *P Natl Acad Sci USA* **116**, 19449-19457 (2019).
17. D. R. Kattnig, J. K. Sowa, I. A. Solov'yov, P. J. Hore, Electron spin relaxation can enhance the performance of a cryptochrome-based magnetic compass sensor. *New Journal of Physics* **18**, 063007 (2016).

18. S. Y. Wong, Y. J. Wei, H. Mouritsen, I. A. Solov'yov, P. J. Hore, Cryptochrome magnetoreception: four tryptophans could be better than three. *Journal of the Royal Society Interface* **18** (2021).
19. Y. T. Kao *et al.*, Ultrafast dynamics and anionic active states of the flavin cofactor in cryptochrome and photolyase. *J Am Chem Soc* **130**, 7695-7701 (2008).
20. A. Lukacs, A. P. M. Eker, M. Byrdin, K. Brettel, M. H. Vos, Electron Hopping through the 15 angstrom Triple Tryptophan Molecular Wire in DNA Photolyase Occurs within 30 ps. *J Am Chem Soc* **130**, 14394-+ (2008).
21. B. Paulus *et al.*, Spectroscopic characterization of radicals and radical pairs in fruit fly cryptochrome-protonated and nonprotonated flavin radical-states. *Febs Journal* **282**, 3175-3189 (2015).
22. K. Schulten, C. E. Swenberg, A. Weller, Biomagnetic sensory mechanism based on magnetic-field modulated coherent electron-spin motion. *Zeitschrift Fur Physikalische Chemie-Frankfurt* **111**, 1-5 (1978).
23. T. Ritz, S. Adem, K. Schulten, A model for photoreceptor-based magnetoreception in birds. *Biophys J* **78**, 707-718 (2000).
24. I. A. Solov'yov, D. E. Chandler, K. Schulten, Exploring the possibilities for radical pair effects in cryptochrome. *Plant Signaling & Behavior* **3**, 676-677 (2008).
25. D. P. Zhong, A. H. Zewail, Femtosecond dynamics of flavoproteins: Charge separation and recombination in riboflavine (vitamin B-2)-binding protein and in glucose oxidase enzyme. *P Natl Acad Sci USA* **98**, 11867-11872 (2001).
26. Y. T. Kao *et al.*, Ultrafast dynamics of flavins in five redox states. *J Am Chem Soc* **130**, 13132-13139 (2008).
27. D. Immeln, A. Weigel, T. Kottke, J. L. P. Lustres, Primary Events in the Blue Light Sensor Plant Cryptochrome: Intraprotein Electron and Proton Transfer Revealed by Femtosecond Spectroscopy. *J Am Chem Soc* **134**, 12536-12546 (2012).
28. R. J. Kutta, N. Archipowa, N. S. Scrutton, The sacrificial inactivation of the blue-light photosensor cryptochrome from Drosophila melanogaster. *Phys Chem Chem Phys* **20**, 28767-28776 (2018).
29. H. Chosrowjan, S. Taniguchi, N. Mataga, F. Tanaka, A. J. W. G. Visser, The stacked flavin adenine dinucleotide conformation in water is fluorescent on picosecond timescale. *Chem Phys Lett* **378**, 354-358 (2003).
30. J. Brazard *et al.*, New Insights into the Ultrafast Photophysics of Oxidized and Reduced FAD ins Solution. *Journal of Physical Chemistry A* **115**, 3251-3262 (2011).
31. A. Sancar, Structure and Function of DNA Photolyase and Cryptochrome Blue-Light Photoreceptors. *Chemical Reviews* **103**, 2203-2238 (2003).
32. J. Brazard *et al.*, Primary Photoprocesses Involved in the Sensory Protein for the Photophobic Response of Blepharisma japonicum. *J Phys Chem B* **112**, 15182-15194 (2008).
33. J.-P. Bouly *et al.*, Cryptochrome Blue Light Photoreceptors Are Activated through Interconversion of Flavin Redox States *[*]. *J Biol Chem* **282**, 9383-9391 (2007).
34. J. Shirdel, P. Zirak, A. Penzkofer, H. Breitkreuz, E. Wolf, Absorption and fluorescence spectroscopic characterisation of the circadian blue-light photoreceptor cryptochrome from Drosophila melanogaster (dCry). *Chem Phys* **352**, 35-47 (2008).
35. Z. Y. Liu *et al.*, Determining complete electron flow in the cofactor photoreduction of oxidized photolyase. *P Natl Acad Sci USA* **110**, 12966-12971 (2013).
36. R. Martin *et al.*, Ultrafast flavin photoreduction in an oxidized animal (6-4) photolyase through an unconventional tryptophan tetrad. *Phys Chem Chem Phys* **19**, 24493-24504 (2017).
37. I. A. Solov'yov, T. Domratcheva, A. R. Moughal Shahi, K. Schulten, Decrypting Cryptochrome: Revealing the Molecular Identity of the Photoactivation Reaction. *J Am Chem Soc* **134**, 18046-18052 (2012).

38. G. Lüdemann, I. A. Solov'yov, T. Kubař, M. Elstner, Solvent Driving Force Ensures Fast Formation of a Persistent and Well-Separated Radical Pair in Plant Cryptochrome. *J Am Chem Soc* **137**, 1147-1156 (2015).
39. S. Y. Qin *et al.*, A magnetic protein biocompass. *Nature Materials* **15**, 217-+ (2016).
40. C. C. Moser, C. C. Page, R. Farid, P. L. Dutton, Biological electron transfer. *Journal of Bioenergetics and Biomembranes* **27**, 263-274 (1995).
41. C. C. Moser, J. M. Keske, K. Warncke, R. S. Farid, P. L. Dutton, Nature of biological electron transfer. *Nature* **355**, 796-802 (1992).
42. C. C. Page, C. C. Moser, X. Chen, P. L. Dutton, Natural engineering principles of electron tunnelling in biological oxidation–reduction. *Nature* **402**, 47-52 (1999).
43. D. Mims, J. Herpich, N. N. Lukzen, U. E. Steiner, C. Lambert, Readout of spin quantum beats in a charge-separated radical pair by pump-push spectroscopy. *Science* **374**, 1470-1474 (2021).
44. P. Muller, J. Yamamoto, R. Martin, S. Iwai, K. Brettel, Discovery and functional analysis of a 4th electron-transferring tryptophan conserved exclusively in animal cryptochromes and (6-4) photolyases. *Chem Commun* **51**, 15502-15505 (2015).
45. F. Lacombat *et al.*, Ultrafast Oxidation of a Tyrosine by Proton-Coupled Electron Transfer Promotes Light Activation of an Animal-like Cryptochrome. *J Am Chem Soc* **141**, 13394-13409 (2019).
46. A. Gunther *et al.*, Double-Cone Localization and Seasonal Expression Pattern Suggest a Role in Magnetoreception for European Robin Cryptochrome 4. *Current Biology* **28**, 211-+ (2018).
47. K. Schwinn, N. Ferre, M. Huix-Rotllant, UV-visible absorption spectrum of FAD and its reduced forms embedded in a cryptochrome protein. *Phys Chem Chem Phys* **22**, 12447-12455 (2020).
48. P. Mondal, K. Schwinn, M. Huix-Rotllant, Impact of the redox state of flavin chromophores on the UV-vis spectra, redox and acidity constants and electron affinities. *J Photoch Photobio A* **387** (2020).
49. I. H. M. van Stokkum, D. S. Larsen, R. van Grondelle, Global and target analysis of time-resolved spectra. *Biochimica Et Biophysica Acta-Bioenergetics* **1657**, 82-104 (2004).
50. S. Solar, N. Getoff, P. S. Surdhar, D. A. Armstrong, A. Singh, Oxidation of Tryptophan and N-Methylindole by N3., Br2-, and (Scn)2- Radicals in Light-Water and Heavy-Water Solutions - a Pulse-Radiolysis Study. *J Phys Chem-Us* **95**, 3639-3643 (1991).
51. J. J. Hopfield, Electron Transfer Between Biological Molecules by Thermally Activated Tunneling. *Proceedings of the National Academy of Sciences* **71**, 3640-3644 (1974).
52. R. A. Marcus, On the Theory of Chemiluminescent Electron-Transfer Reactions. *The Journal of Chemical Physics* **43**, 2654-2657 (1965).
53. J. Blumberger, Recent Advances in the Theory and Molecular Simulation of Biological Electron Transfer Reactions. *Chemical Reviews* **115**, 11191-11238 (2015).
54. D. Brida, C. Manzoni, G. Cerullo, Phase-locked pulses for two-dimensional spectroscopy by a birefringent delay line. *Opt Lett* **37**, 3027-3029 (2012).
55. D. C. Lunemann *et al.*, Distinguishing between coherent and incoherent signals in excitation-emission spectroscopy. *Opt Express* **29**, 24326-24337 (2021).
56. M. Rabe, Spectram: A MATLAB® and GNU Octave Toolbox for Transition Model Guided Deconvolution of Dynamic Spectroscopic Data. *Journal of Open Research Software* **8**, 13 (2020).
57. D. R. Kattnig, C. Nielsen, I. A. Solov'yov, Molecular dynamics simulations disclose early stages of the photo-activation of cryptochrome 4. *New Journal of Physics* **20**, 083018 (2018).
58. E. Sjulstok, G. Ludemann, T. Kubar, M. Elstner, I. A. Solov'yov, Molecular Insights into Variable Electron Transfer in Amphibian Cryptochrome. *Biophys J* **114**, 2563-2572 (2018).
59. V. Korol *et al.*, Introducing VIKING: A Novel Online Platform for Multiscale Modeling. *ACS Omega* **5**, 1254-1260 (2020).

**Figures and Tables**

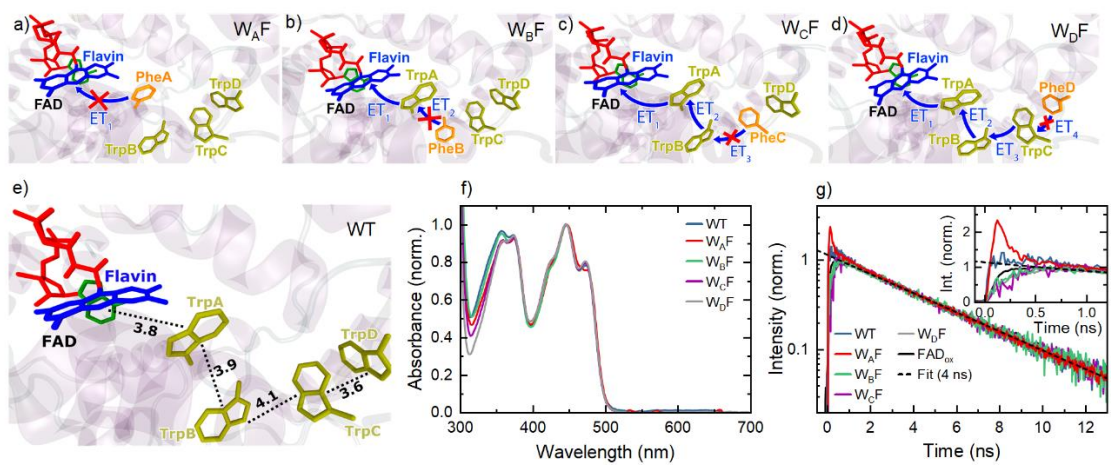

**Figure 1.** Visualization of the principal electron transfer route in the *Er*Cry4 protein and the effect of single amino acid mutations on blocking the electron transfer process at different stages. (a)-(d) Electron transfers in *Er*Cry4 mutants involving Trp → Phe substitutions with W395 (Trp$_A$), W372 (Trp$_B$), W318 (Trp$_C$) and W369 (Trp$_D$). The arrows illustrate the electron transfer paths while the crossed arrows represent the electron transfers that are blocked, as suggested by the experimental data in this study. (e) The edge-to-edge distances between the FAD chromophore and the important tryptophans involved for the wild type (WT) *Er*Cry4 protein. (f) Normalized linear absorption spectra of fully oxidized wild type *Er*Cry4 and its four mutants. The pronounced vibronic substructure around 450 nm indicates FAD binding to the protein. (g) Transient PL decay curves of wild type *Er*Cry4 and its mutants compared to that of free FAD$_{ox}$ in buffer solution and fit to a monoexponential, 4 ns decay (solid lines). Only mutant W$_A$F shows an initial fast component originating from weak PL emission of protein bound FAD (inset).

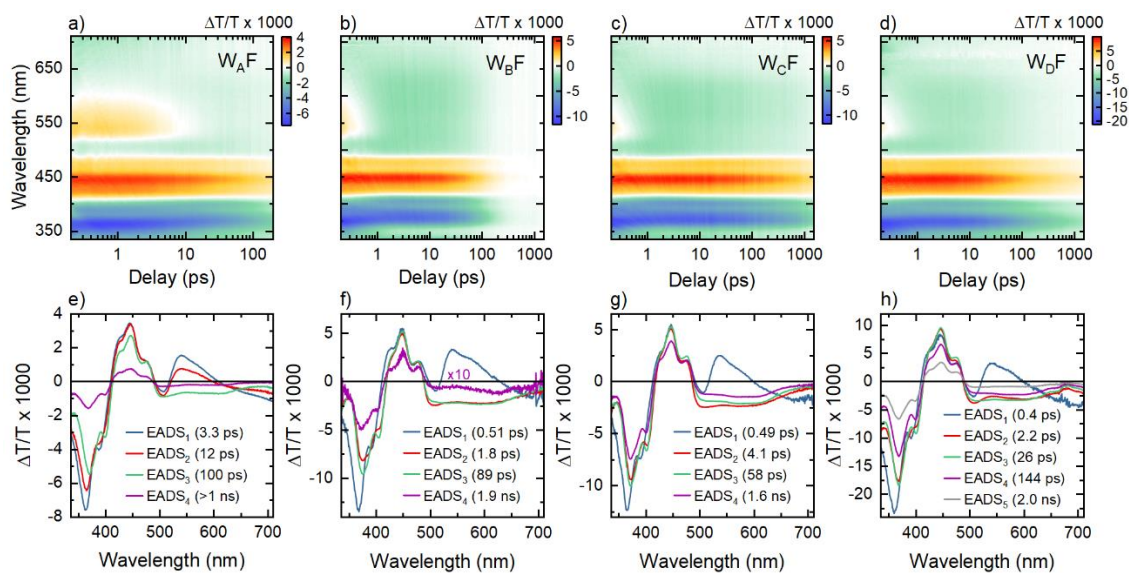

**Figure 2.** Transient absorption spectroscopy of *Er*Cry4 mutants. (a)-(d) Differential transmission $\Delta T/T$ spectra of the mutants $W_AF$ to $W_DF$ for delays between 0.2 ps and 2 ns (logarithmic time axis). (e)-(h) Corresponding evolution associated difference spectra (EADS) for each mutant resulting from a global data analysis. (e) The $W_AF$ mutant does not show a sub-ps decay component since the electron transfer from Trp$_A$ to FAD is suppressed. (f-h) The 0.5 ps decay components in the EADS spectra of $W_BF$ - $W_DF$ reflects this primary electron transfer from Trp$_A$ to FAD. Each increase in distance between Phe mutation and chromophore ($W_BF \rightarrow W_DF$) results in one additional decay component in the EADS spectra, giving evidence for the second and third electron transfer step in the sequential electron transfer chain in *Er*Cry4. A 2-4 ps decay component in all 4 mutants is assigned to vibrational cooling of the optically created radical pairs. The slowest component in all mutants reflect radical pair recombination on a ns time scale.

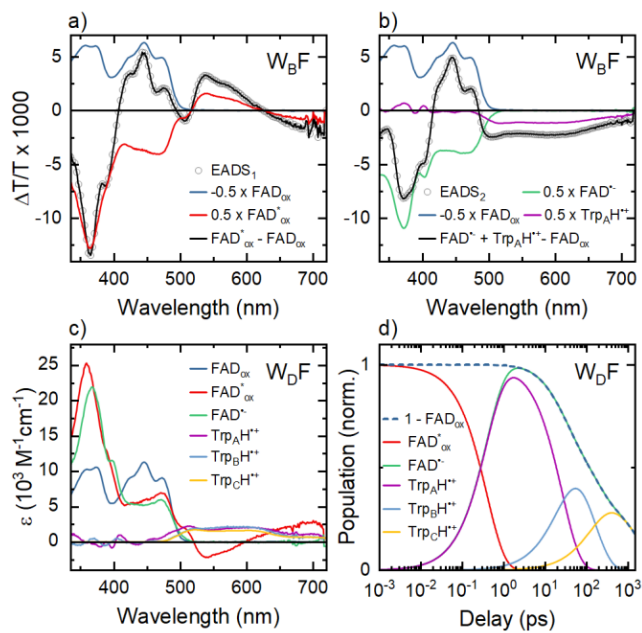

**Figure 3.** Analysis of the different transmission $\Delta T/T$ spectra of *Er*Cry4 mutants. a) EADS spectrum EADS$_1$ of mutant W$_B$F with 0.51 ps decay time (open circles) and decomposition into the $\Delta T/T$ spectra of FAD$_{ox}$ and FAD$_{ox}^*$ (solid lines). b) EADS spectrum EADS$_2$ of mutant W$_B$F with 1.8 ps decay time (open circles) and decomposition into the $\Delta T/T$ spectra of FAD$_{ox}$, FAD$^{\bullet-}$ and Trp$_A$H$^{\bullet+}$ (solid lines). c) Differential transmission $\Delta T/T$ spectra of the different neutral and charged radical components in the *Er*Cry4 protein as deduced from the analysis of mutant W$_D$F. (d) Population dynamics of the different neutral and charged radical states in mutant W$_D$F.

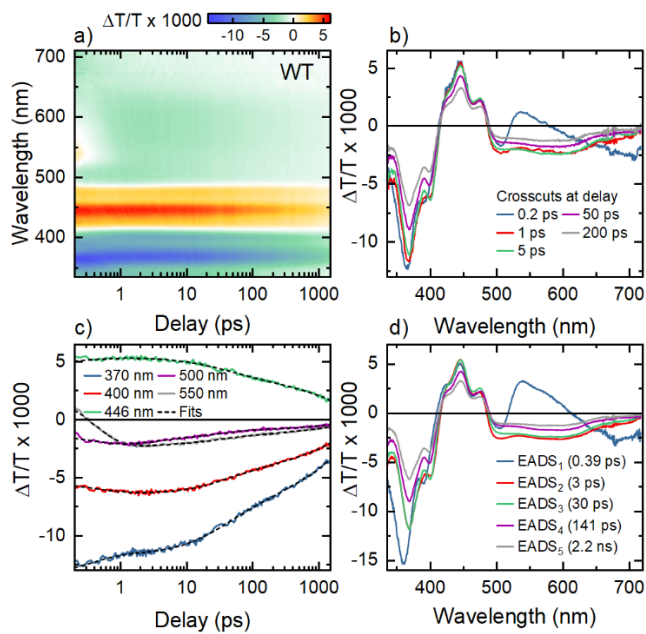

**Figure 4.** Transient absorption spectroscopy of wild type *Er*Cry4 proteins. (a) Differential transmission $\Delta T/T$ spectra of the wild type protein for delays been 0.2 ps and 1.5 ns (logarithmic time axis). b) $\Delta T/T$ spectra at selected delays. c) Dynamics of the $\Delta T/T$ spectra at selected wavelengths. The results of the global analysis are added as black dashed lines. d) EADS spectra as obtained from the global data analysis. The resulting spectra are very similar to those of mutant $W_D F$. Based on the analysis of the mutant spectra, we assign the components with decay times of 0.39 ps, 30 ps, and 141 ps to the first, second and third sequential electron transfer across the tryptophan chain of the protein. The 3 ps decay component is associated with vibrational cooling of the optically generated radical pairs, while the 2.2 ns component reflects radical pair recombination.

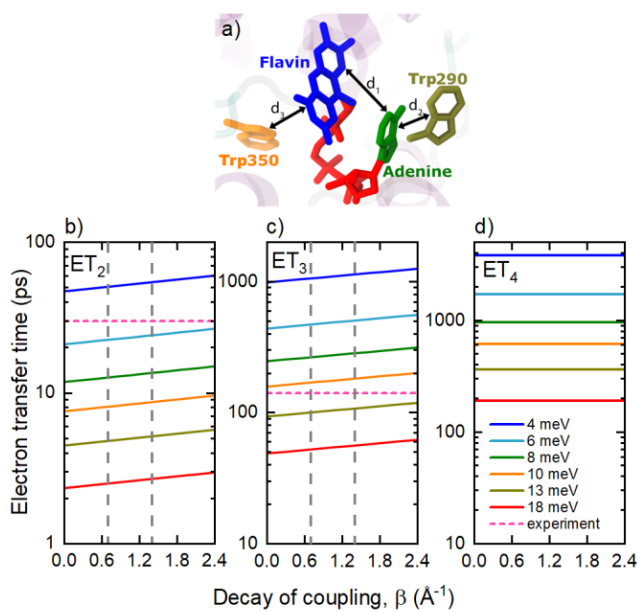

**Figure 5.** (a) Flavin adenine dinucleotide (FAD) cofactor with the indicated internal edge-to-edge distances between the flavin and adenine ($d_1$ = 0.36 nm) cofactors, the edge-to-edge distance between adenine and Trp290 residue ($d_2$ = 0.39 nm) and the edge-to-edge distance between the flavin and the Trp350 residue ($d_3$ = 0.73 nm), see Fig. S9. (b-d) Characteristic electron transfer times, required for electron transfers 2, 3 and 4 respectively (see Fig. 1). The solid lines represent the predictions of the Marcus-Hopfield theory for different electron couplings $V_0$, Eq. (1), of 6, 8, 10, 13, 18 and 23 meV which are typical values for ET in proteins. The dashed lines show the experimental values deduced from the analysis of Fig. 4.

**Table 1.** Electron transfer times and yields for the investigated mutants and the wild type protein.

| Sample | $ET_1$ ($\eta_1$) | $ET_2$ ($\eta_2$) | $ET_3$ ($\eta_3$) |
|---|---|---|---|
| $W_BF$ | 0.51 ps (100%) | | |
| $W_CF$ | 0.49 ps (100%) | 58 ps (77%) | |
| $W_DF$ | 0.40 ps (100%) | 26 ps (71%) | 144 ps (52%) |
| WT | 0.39 ps (100%) | 30 ps (78%) | 141 ps (77%) |

# Supplementary Information for

Tracking the Electron Transfer Cascade in European Robin Cryptochrome 4 Mutants

Daniel Timmer[a], Daniel C. Lünemann[a], Anitta R. Thomas[a], Anders Frederiksen[a], Jingjing Xu[b], Rabea Bartölke[b], Jessica Schmidt[b], Antonietta De Sio[a], Ilia A. Solov'yov[a,c], Henrik Mouritsen[b,c], Christoph Lienau[a,c].

[a] Institut für Physik, Carl von Ossietzky Universität, 26129 Oldenburg, Germany
[b] Institut für Biologie und Umweltwissenschaften, Carl von Ossietzky Universität, 26129 Oldenburg, Germany
[c] Research Centre for Neurosensory Science, Carl von Ossietzky Universität, 26111 Oldenburg, Germany

*Corresponding author(s): Christoph Lienau.

**Email:** christoph.lienau@uni-oldenburg.de

**Photoluminescence excitation experiments**

Photoluminescence (PL) excitation experiments are performed by use of an interferometric time correlated single photon counting (TCSPC) setup (1). Here, a 40 MHz white light source (SC400, Fianium) is filtered to a spectral range of 400-680 nm with an average power of 50 µW. It passes a first, passively stable common-path interferometer (Translating Wedge-based Identical pulse eNcoding System, TWINS (2)) and is focused using a microscope objective (0.35 NA) into the sample. The sample solution is kept contained in a Hellma micro-cuvette at 1°C throughout the measurement. Directly before and after the sample a fraction of the incident/transmitted light is split off and measured with a photodiode to monitor the sample absorption over time. A second microscope objective (0.35 NA) placed at 90° with respect to the first one collects the light emitted from the sample that passes through a second TWINS interferometer and is finally focused onto a single-photon avalanche photodiode (PDM, Micro Photon Devices, 100x100 µm² sensor area). Both TWINS are scanned and a TCSPC histogram is recorded by the TCSPC electronics (PicoHarp 300, PicoQuant) at every position, additional to the photodiode signals.

The Fourier transform of the signal along both TWINS position axes yields excitation-emission spectra (3) at each TCSPC time point and the time-integrated excitation-emission spectrum, see Fig. S1a. The interferometric approach also allows us to separate out the coherent scattering contribution (1). A Matlab-based toolbox (4) is used to globally analyze the 3-dimensional dataset, decomposing it into independent sets of excitation and emission spectra and dynamics, see Fig. S1b-c and Fig. 1. The Fourier transform of the photodiode signals yields the spectrum of the exciting laser light and the absorption spectrum of the sample. The excitation-emission maps and excitation spectra are normalized to the white light spectrum to only feature sample properties. To compare the spectra of wild type *Er*Cry4 and the mutants with those of free FAD, a 40 µM solution of molecular FAD in a buffer that contains 1 mM PFC was measured. Excitation and emission spectra of all protein species are indistinguishable from those of free $FAD_{ox}$ and no spectral signatures of protein-bound FAD redox states could be observed (Fig. S1).

This indicates that there are no measurable spectral signatures from protein-bound FAD in the excitation spectra of wild type *Er*Cry4 and its mutants and that essentially all of the emission stems from residual FAD molecules that are not bound to the protein. Only in mutant $W_AF$ we see a faint emission from bound $FAD_{ox}$, more clearly resolved in the time-dependent measurements in Fig. 1g. We have deduced the concentrations of FAD chromophores in the protein samples and that of molecular FAD in solution from absorption measurements. By comparing the signal strength of the PL in the excitation-emission maps of the wild type proteins and the mutants to that of molecular

FAD in buffer solution, we can then estimate the fraction of bound FAD. In all as-prepared samples, the fraction of bound FAD is at least 97%.

**Transient absorption (TA) experiments**

A 10 kHz, 1 mJ regenerative Ti:Sa amplifier (Legend Elite, Coherent) provides pulses with ~25 fs duration centered around 800 nm. The laser is pumping a tunable optical parametric amplifier (TOPAS, Light Conversion) generating 450 nm pulses with a bandwidth of 15 nm. A pair of chirped mirrors (DCM 12, Laser Quantum) is used for chirp compensation of setup dispersion. A home-built Transient Grating Frequency Resolved Optical Gating (TG-FROG) setup (5) provides a measured pulse duration of ~28 fs.

A fraction of the fundamental Ti:Sa light is focused into a 3 mm thick calcium fluoride (CaF$_2$) plate (Eksma Optics), which is continuously moved by two motorized translation stages to avoid photo-induced damage of the crystal over time. The emitted white light spans from ~320-750 nm and is collimated with an off-axis parabolic mirror. Remaining fundamental laser light is attenuated with a colored glass filter (FGS580M, Thorlabs).

Pump and probe beams are both simultaneously chopped (MC2000, Thorlabs) with a 2:1 duty cycle chopper wheel on a single shot basis for data acquisition. This generates a sequence of pump-on/probe-on $S_{\mathrm{pu_{on},\,pr_{on}}}(\lambda, t_W)$, pump-off/probe-on $S_{\mathrm{pu_{on}pr_{off}}}(\lambda)$ and pump-on/probe-off $S_{\mathrm{pu_{off},\,pr_{on}}}(\lambda)$ spectra that is then used for evaluating differential transmission spectra. A fast and sensitive low noise line camera (Aviiva EM4, e2v), mounted to a 150 mm grating monochromator (Acton SP-2150, Princeton Instruments, 150 l/mm grating blazed for 500 nm) records 1000 single shot spectra for each pulse delay $t_W$. From these individual spectra, scattering-corrected differential transmission spectra

$$\frac{\Delta T}{T}(\lambda, t_W) = \frac{S_{\mathrm{pu_{on},\,pr_{on}}}(\lambda, t_W) - S_{\mathrm{pu_{on}pr_{off}}}(\lambda) - S_{\mathrm{pu_{off},\,pr_{on}}}(\lambda)}{S_{\mathrm{pu_{off},\,pr_{on}}}(\lambda)}$$

are calculated and then averaged. The subtraction of the $S_{\mathrm{pu_{on}pr_{off}}}(\lambda)$ spectrum ensures that weak and spurious light scattering contributions on the detector that are induced by the pump pulse are efficiently suppressed. The waiting time delays $t_w$ between pump and probe are changed with a linear translation stage (M531.5IM, Physik Instrumente). A series of delay scans are performed and averaged afterwards.

A temperature-controlled sample chamber, set to 1°C, is continuously moved to minimize effects of photoreduction. The pump and probe beam are focused into the sample under a small angle of ~4° by use of an off-axis parabolic mirror.

TA measurements are performed with 20 nJ pump and ~1 nJ probe pulses at 10 kHz laser repetition rate. The pump pulses are focused to a spot size of ~50x50 μm², giving an excitation fluence of 1 mJ/cm² The polarization of the probe is parallel to the optical table, while the pump has a polarization of 45°. A polarizer mounted on a motorized rotation stage can turn the pump polarization, allowing for parallel and crossed polarization between pump and probe without affecting the pump power.

Experimental data is acquired in crossed and parallel polarization for each scan, from which isotropic differential transmission spectra at magic angle (MA) are calculated as

$$\frac{\Delta T}{T}_{\mathrm{MA}} = \frac{1}{3}\left(\frac{\Delta T}{T}_{\mathrm{parallel}} + 2\frac{\Delta T}{T}_{\mathrm{crossed}}\right)$$

These MA spectra are displayed in the main manuscript and used for further analysis.

**Linearity and sample stability**

To ensure that TA experiments are performed in a linear pump-power regime, the dependence of $\Delta T/T$ signal strength on pump pulse energy was recorded for a wild type sample. The spot sizes of pump and probe at their intersection were set to ~50x50 µm².

Fig. S2a shows that the shape of the $\Delta T/T$ spectra at a fixed delay of 2 ps, normalized to their maximum at 450 nm, does not change with pump energy in the range between 5 and 40 nJ. Also the dynamics at selected delays do not change when varying the pump energy in this range. As seen in Fig. 2c, no deviation from a linear pump-power dependence can be observed. This ensures that all reported TA measurements in this manuscript, recorded at a pump energy of 20 nJ, are well within the linear pump-power regime, not exceeding the $\chi^{(3)}$ regime. To test for possible sample degradation during the nonlinear measurements, we have investigated the time evolution of the differential transmission spectrum of a $W_BF$ mutant at a fixed delay of 2 ps. No visible change in $\Delta T/T$ is seen during the course of the 6.5 h measurement. The experimental conditions were the same as for the data shown in Fig. 2 and 4 of the main manuscript. In Fig. S3b, the dynamics of the $\Delta T/T$ signal are displayed at selected probe wavelengths of 450 nm and 550 nm for all 30 scans shown in Fig. S3a. Also here, we find no change of signal over the span of the measurement time, indicating negligible amounts sample degradation during the 6.5 h measurement period, longer than any measurement reported in the main manuscript.

**PFC oxidizer**

In all experiments that are shown in the main manuscript, potassium ferricyanide (PFC) has been added as an oxidizing agent to counteract photoreduction of the sample during the measurement (6). Figure S4a shows the linear absorption spectrum of 1 mM PFC in Tris buffer (20% glycerol, 200 mM NaCl, pH 8.0). The main absorption band around ~420 nm has some overlap with the 450 nm pump pulse spectrum.

The addition of ~1.5 mM PFC during sample preparation will oxidize a partially reduced sample. This oxidation may partially or even fully deplete the PFC concentration. In the latter case, additional PFC has been added until the sample shows no residual absorption of reduced FAD. A reduced sample shows a broad FADH absorption band in the range of 500-700 nm, in contrast to a fully reduced sample only containing $FAD_{ox}$. The amount of PFC remaining in the now fully oxidized sample is then determined using linear absorption measurements. The remaining amount of PFC, which counteracts photo-reduction during the experiments is found to range from ~1.7-2.6 mM in the investigated samples at the start of the measurement.

A TA measurement of 1 mM PFC in buffer under the same experimental conditions as the *Er*Cry4 measurements is shown in Figure S4b. Here, only a very weak differential transmission signal on the order of 0.01% can be observed, decaying within a few ps. This makes possible PFC contributions to the nonlinear signal of the protein measurements negligible.

**FAD in buffer**

A TA measurement of a 200 µM solution of molecular FAD dissolved in a buffer containing 1 mM PFC is shown in Fig. S5. The sample was measured under the same experimental conditions as used for the *Er*Cry4 measurements. A global analysis of the data results in four DADS spectra that are displayed in Fig. S5d. The spectral shapes of these DADS spectra and their associated decay times closely follow those reported by Brazard et al. (7). Here, the long-lived 3.2 ns component (3.0 ns in Ref. (7)) can be assigned to the relaxation of the open conformer of optically excited FAD, in which the distance between the electron-donating adenine moiety and the electron-accepting isoalloxazine group is so large that an intramolecular electron transfer is efficiently suppressed. The DADS spectrum with a much shorter decay time of 6.6 ps (5.4 ps in Ref. (7)) is assigned to monitor the intramolecular electron transfer between adenine and isoalloxazine in the closed conformer of FAD. Additional DADS spectra with decay times of 1.7 ps and 35 ps (1 ps and 31 ps in Ref. (7), respectively) are attributed to solvation dynamics and to the decay of the FAD dimer, respectively. No effect of PFC on dynamics and spectral shapes can be observed.

**Data preparation and analysis**

For each *Er*Cry4 measurement, a measurement of the differential transmission signal of a plain buffer solution was performed. This measurement shows, within +/- 100 fs around time zero and at all probe wavelengths, a well-known and strong coherent scattering contribution that arises from the cross-phase modulation between the chirped few-ps probe pulse and the short pump pulse (8). This coherent scattering signal can be employed to measure the wavelength-dependent zero point of the delay, $t_W = 0$. For this, we have used the method described in (9) to fit the solvent signal. Based on this fit, the origin of the time delay axis was corrected for each TA measurement. Afterwards, the coherent solvent contribution, as obtained from the reference measurement, has been subtracted from the protein measurement. To avoid any spurious, residual solvent contribution in the measurements discussed in this manuscript, we have discarded the TA data for delays below 200 fs from the data analysis.

To investigate the *Er*Cry4 wild type and mutant datasets, a global analysis was performed by use of a Matlab-based toolbox (10). To obtain decay associated difference spectra (DADS), only monoexponential decays were used as mathematical model functions. Each DADS component, $DADS_i(\lambda), i = 1, \ldots, n$, represents the spectral amplitude of the signal contribution that is decaying with a time constant $\tau_i$. In this way, the global fit will reproduce the complete, experimentally measured data set by a sum over all $DADS_i(\lambda)$ spectra multiplied by an exponential decay with decay time $\tau_i$

$$\frac{\Delta T}{T}(\lambda, t_W) = \sum_{i=1}^{n} DADS_i(\lambda)\, e^{-\frac{t_W}{\tau_i}}$$

The *i*-th term of this sum, $DADS_i \exp(-t_W/\tau_i)$, is the time-dependent differential transmission spectrum associated with the relaxation component *i*. This representation in terms of independent components with monotonically increasing decay times accurately describes the dynamics expected from a sequential rate equation model.

The DADS spectra that result from the global analysis for all mutants and the wild type are shown in Fig. S7. The datasets were modeled using the smallest number of decays necessary to fully describe the dynamics of the data at all wavelengths. The correctness of this number was judged by the appearance of temporally and spectrally flat and unstructured residuals. For wild type *Er*Cry4 and the $W_DF$ mutant five and for the $W_AF$, $W_BF$ and $W_CF$ mutants four exponential decays were needed to represent the measured data. Figure S6e-h shows the fits resulting from the global analysis compared to the experimental data dynamics at selected wavelengths. A convincing agreement between fit and data validates the DADS obtained by the global analysis.

These DADS spectra may either reflect the exponential decay or the rise of a certain spectral component in the data. As such, their physical interpretation may sometimes be challenging when spectral components that rise in time overlap with decaying components. This is indeed the case for the present samples where some spectral signatures of the rise in radical pair concentration overlap with decaying signals from oxidized flavin. The physical interpretation of the spectra may be facilitated by transforming them into evolution-associated difference spectra (EADS) that are defined (11) as

$$EADS_k(\lambda) = \sum_{i=k}^{n} DADS_i(\lambda).$$

The first spectrum, $EADS_1$, represents the sum over all DADS spectra and, thus, the differential transmission spectrum at delay zero, $\Delta T/T(\lambda, t_W = 0)$. The *k*-th spectrum, $EADS_k, k > 1$, then gives a differential transmission spectrum which is the difference between $\Delta T/T(t_W = 0)$ and the sum of the first $k - 1$ DADS spectra. Provided that the decay constants are sufficiently different, this approximately represents the shape of the differential transmission at a delay shortly after these $k - 1$ relaxation steps have been completed.

The EADS spectra are then used to extract spectra for different electronic states of FAD ($FAD_{ox}$, $FAD^*_{ox}$ and $FAD^{•-}$), as well as for the different tryptophans ($Trp_A$, $Trp_B$, $Trp_C$). The absorption spectra of the excited flavin, $A_{FAD^*_{ox}}$, and the third radical pair $A_{RP3}$ were obtained by removing the measured absorbance $A_{FAD_{ox}}$ from $EADS_1$ and $EADS_5$, which were converted from differential transmission $\Delta T/T$ to differential absorbance $\Delta A$ using

$$A_{FAD^*_{ox}}(\lambda) = \log_{10}(1 - EADS_1(\lambda)) + nA_{FAD_{ox}}, \text{ and}$$

$$A_{RP3}(\lambda) = A_{FAD^{•-}}(\lambda) + A_{Trp_CH^{•+}}(\lambda) = \log_{10}(1 - EADS_5(\lambda)) + nA_{FAD_{ox}}.$$

The scaling factor $n$ is is chosen as 0.025 such that the obtained spectra are flat and unstructured at the peaks around 450 nm, similar to procedures described in (12, 13).

Literature spectra of TrpH$^+$ show mainly absorption above ~500 nm and below ~380 nm (14), while FAD$^{•-}$ only absorbs at wavelengths below ~500 nm (15, 16). Thus, to disentangle the absorbance spectrum $A_{RP3}$ into its components $A_{FAD^{•-}}$ and $A_{Trp_CH^{•+}}$, a sum of 11 Gaussian spectra is fitted to $A_{RP3}$, where 4 Gaussians represent the plateau for wavelengths above ~500 nm, taken as the Trp$_C$H$^{•+}$ absorbance. After subtracting $A_{Trp_CH^{•+}}$ from $A_{RP3}$, the remaining FAD$^{•-}$ absorbance is used to extract the Trp$_B$H$^{•+}$ and Trp$_A$H$^{•+}$ spectra from $EADS_4$ and $EADS_2$, respectively.

Absorbance is converted to molar extinction $\varepsilon$ using the experimentally measured FAD$_{ox}$ absorbance and a reference value of 11300 M$^{-1}$cm$^{-1}$ for the molar extinction of FAD$_{ox}$ at a wavelength of 450 nm (7, 11, 12). This approach results in the species spectra depicted in Fig. 3c, which are very similar to those reported by Kutta et al. (12).

**SI References**

1. D. C. Lünemann *et al.*, Distinguishing between coherent and incoherent signals in excitation-emission spectroscopy. *Opt Express* **29**, 24326-24337 (2021).

2. D. Brida, C. Manzoni, G. Cerullo, Phase-locked pulses for two-dimensional spectroscopy by a birefringent delay line. *Opt. Lett.* **37**, 3027-3029 (2012).

3. A. Perri *et al.*, Hyperspectral imaging with a TWINS birefringent interferometer. *Opt Express* **27**, 15956-15967 (2019).

4. C. M. Andersen, R. Bro, Practical aspects of PARAFAC modeling of fluorescence excitation-emission data. *J Chemometr* **17**, 200-215 (2003).

5. J. N. Sweetser, D. N. Fittinghoff, R. Trebino, Transient-grating frequency-resolved optical gating. *Opt Lett* **22**, 519-521 (1997).

6. J. J. Xu *et al.*, Magnetic sensitivity of cryptochrome 4 from a migratory songbird. *Nature* **594**, 535-+ (2021).

7. J. Brazard *et al.*, New Insights into the Ultrafast Photophysics of Oxidized and Reduced FAD ins Solution. *J Phys Chem A* **115**, 3251-3262 (2011).

8. K. Ekvall *et al.*, Cross phase modulation artifact in liquid phase transient absorption spectroscopy. *J Appl Phys* **87**, 2340-2352 (2000).

9. B. Baudisch (2018) Time resolved broadband spectroscopy from UV to NIR.

10. M. Rabe, Spectram: A MATLAB® and GNU Octave Toolbox for Transition Model Guided Deconvolution of Dynamic Spectroscopic Data. *Journal of Open Research Software* **8**, 13 (2020).

11. F. Lacombat *et al.*, Ultrafast Oxidation of a Tyrosine by Proton-Coupled Electron Transfer Promotes Light Activation of an Animal-like Cryptochrome. *J Am Chem Soc* **141**, 13394-13409 (2019).

12. R. J. Kutta, N. Archipowa, N. S. Scrutton, The sacrificial inactivation of the blue-light photosensor cryptochrome from Drosophila melanogaster. *Phys Chem Chem Phys* **20**, 28767-28776 (2018).

13. R. J. Kutta, N. Archipowa, L. O. Johannissen, A. R. Jones, N. S. Scrutton, Vertebrate Cryptochromes are Vestigial Flavoproteins. *Sci Rep-Uk* **7**, 44906 (2017).

14. S. Solar, N. Getoff, P. S. Surdhar, D. A. Armstrong, A. Singh, Oxidation of Tryptophan and N-Methylindole by N3., Br2-, and (Scn)2- Radicals in Light-Water and Heavy-Water Solutions - a Pulse-Radiolysis Study. *J Phys Chem-Us* **95**, 3639-3643 (1991).

15. K. Schwinn, N. Ferre, M. Huix-Rotllant, UV-visible absorption spectrum of FAD and its reduced forms embedded in a cryptochrome protein. *Phys Chem Chem Phys* **22**, 12447-12455 (2020).

16. Y. T. Kao *et al.*, Ultrafast dynamics of flavins in five redox states. *J Am Chem Soc* **130**, 13132-13139 (2008).

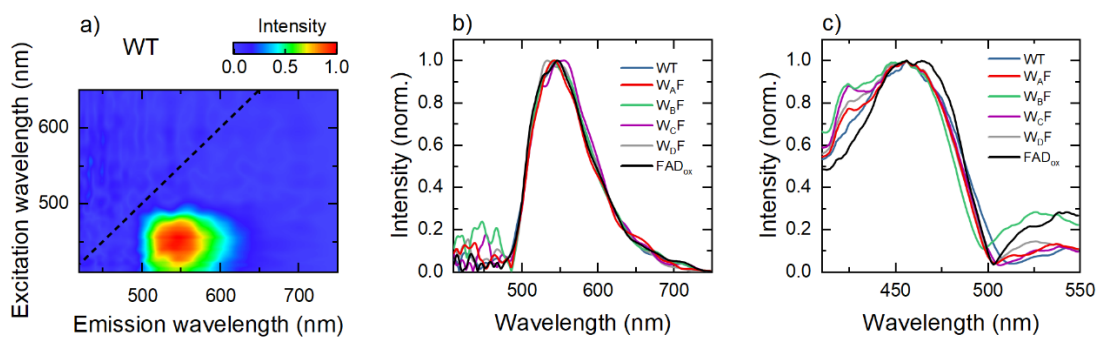

**Figure S1.** PL spectra of wild type *Er*Cry4 and mutants. a) Excitation-emission map of wild type *Er*Cry4 showing the excitation of a small amount of free FAD that is not bound to the protein. The emission from protein-bound $FAD_{ox}$ is too weak to be seen in this map. b) Emission and c) excitation spectra extracted from the PL maps of wild type *Er*Cry4 (blue lines) and mutants. The lack of a vibronic substructure in the excitation spectra around 450 nm is the marker that the emitting species are not bound to the protein. The shape of the spectra matches that of free $FAD_{ox}$ in buffer (black lines).

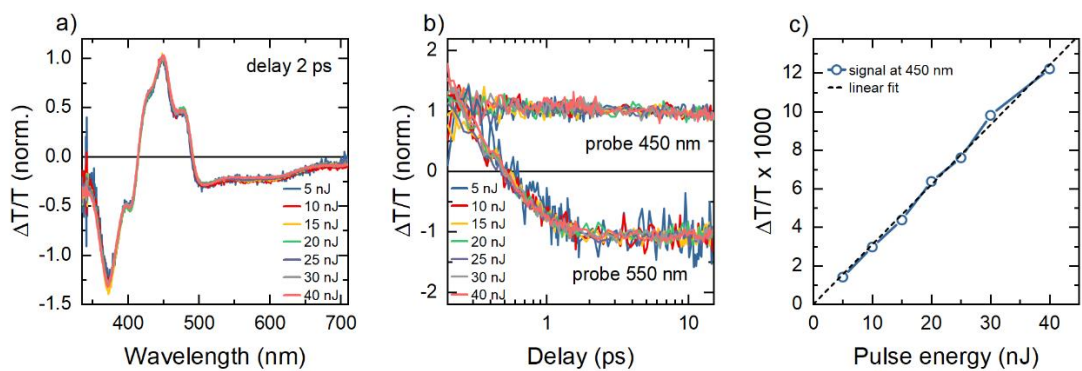

**Figure S2.** Effect of the pump pulse energy on $\Delta T/T$ signals for wild type *Er*Cry4 protein. a) Normalized differential spectra at a delay of 2 ps and b) normalized dynamics at 450 nm and 550 nm recorded for pump pulse energy ranging from 5 nJ to 40 nJ. c) Differential transmission signal averaged over the delays from 5 ps to 15 ps at 450 nm as a function of the pump pulse energy (blue circles) together with a linear fit through the origin. This linear relationship of signal strength and pump pulse energy shows that the experiments are performed well within the linear pump power regime.

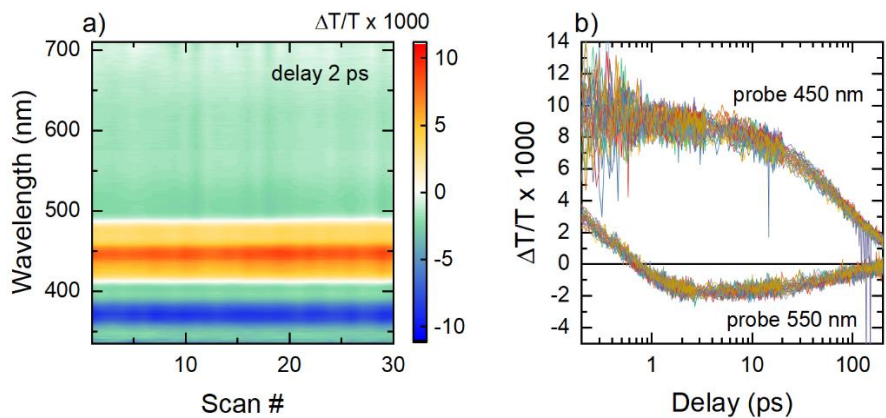

**Figure S3.** Stability of the differential transmission spectra over 30 delay scans, for a measurement of the W$_B$F mutant, lasting ~6.5 h. The measurement is performed for parallel pump/probe polarization using 20 nJ pump pulses at 450 nm. a) Raw differential transmission spectra at a delay of 2 ps for all 30 scans show no visible change during the entire measurement. b) The dynamics of all scans at 450 nm and 550 nm further indicate no significant sample degradation during the measurement. Fluctuations at 450 nm within the first ps arise from interferences between the probe and scattered pump pulse, remaining after scattering correction.

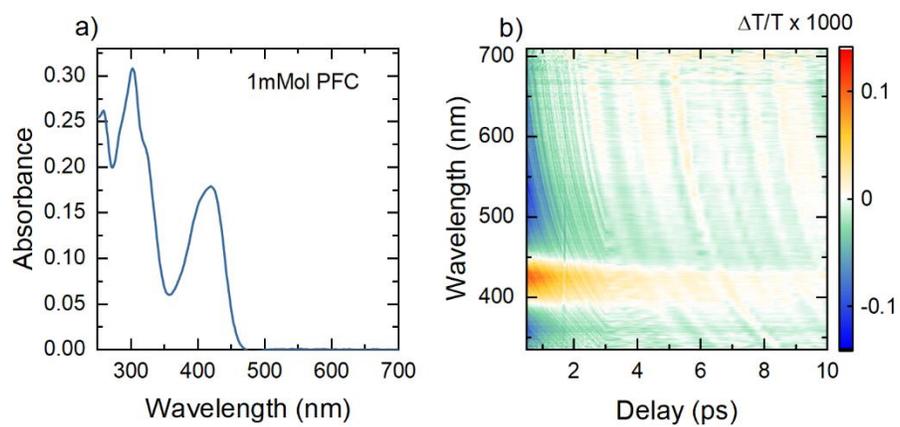

**Figure S4.** Linear absorption and nonlinear differential transmission spectra of PFC oxidizer. a) Linear absorption spectrum of 1 mM PFC in Tris buffer. b) Probe-chirp-corrected $\Delta T/T$ map of 1 mM PFC in buffer recorded under the same experimental conditions as for the *Er*Cry4 measurements. Note that the differential transmission signal is $< 10^{-4}$.

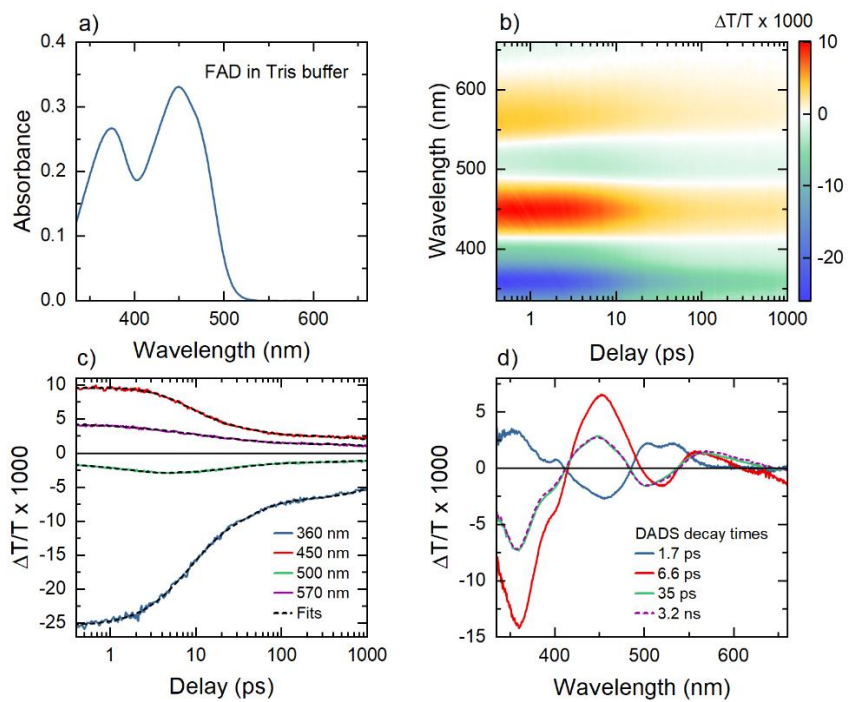

**Figure S5.** Linear absorption and nonlinear differential transmission of molecular FAD in buffer solution. a) Linear absorption spectrum of 0.2 mM FAD in Tris buffer (20% glycerol, 200 mM NaCl, pH 8.0). b) Probe-chirp-corrected $\Delta T/T$ map of 200 µM FAD in Tris buffer containing 2 mM PFC, measured under the same experimental conditions as the *Er*Cry4 measurements. c) Dynamics of the differential transmission signal for selected wavelengths. The result of the global analysis of the data is superimposed as dashed black lines. d) The four DADS spectra that are obtained from a global analysis of the data and their associated decay times. Spectra and decay times are very similar to those reported in Ref. (7) for a sample with a 0.23 mM concentration. No effects of the PFC oxidizer on the dynamics and on the spectra can be observed.

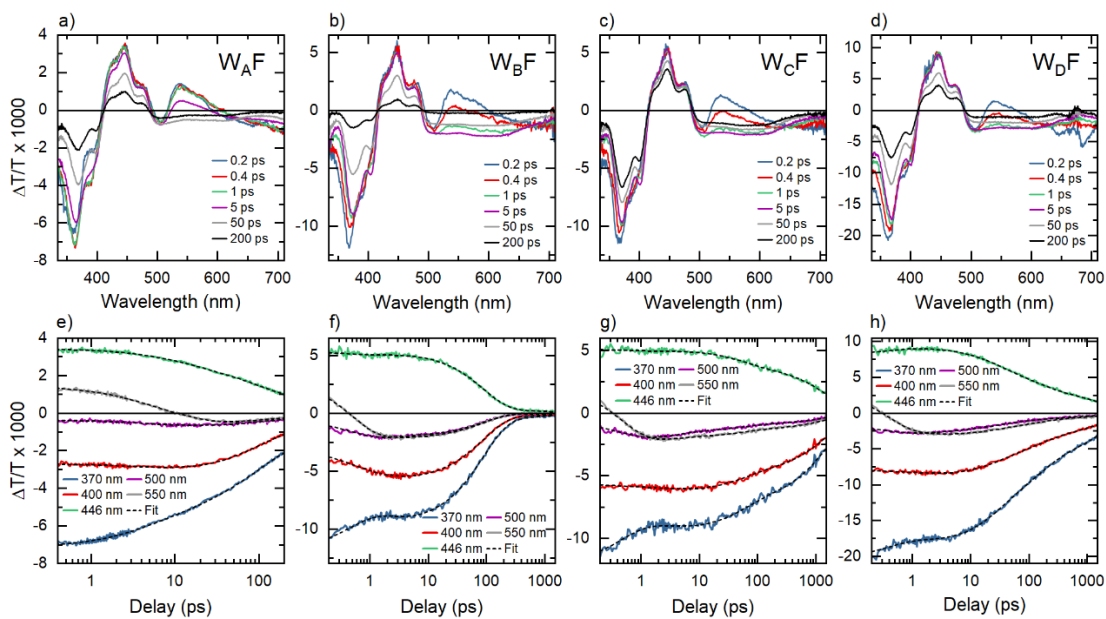

**Figure S6.** a)-d): Differential transmission spectra of all four investigated *Er*Cry4 mutants at selected delays. The spectra are cross-sections through the data presented in Fig. 2 of the main manuscript. e)-h): Dynamics at selected wavelengths for the four mutants (solid lines). The fits obtained from the global analysis (black dashed lines) show excellent agreement with the experimental data.

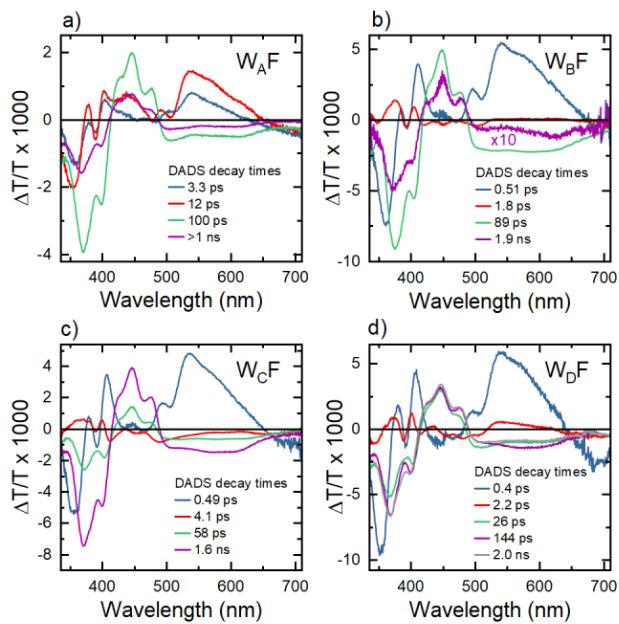

**Figure S7.** DADS spectra resulting from a global analysis of the differential transmission maps recorded for the four *Er*Cry4 mutants (Fig. 2 of the main manuscript). The global analysis results in a minimal number of physically meaningful decay components with different differential transmission spectra and associated decay times that are depicted in the individual subpanels. These DADS spectra are then used to calculate the EADS spectra of the four mutants that are shown in Figs. 2e-h of the main manuscript.

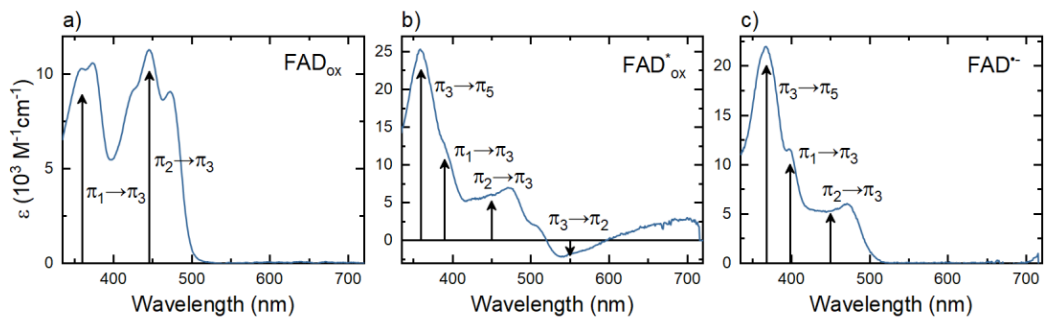

**Figure S8.** Spectral assignment of the differential transmission spectra of a) $FAD_{ox}$, b) $FAD^*_{ox}$ and c) $FAD^{•-}$ as obtained from the $W_DF$ dataset. The assignment is based on quantum-chemical calculations recently reported by Schwinn et al. (15). The corresponding intramolecular transitions for $FAD_{ox}$ and $FAD^{•-}$ are marked with black arrows. The molecular orbitals that are involved in the $FAD^*_{ox}$ absorption are assumed to be the same as those for $FAD^{•-}$. An additional stimulated emission peak ($\pi_3 \rightarrow \pi_2$) appears in $FAD^*_{ox}$ at around 550 nm.

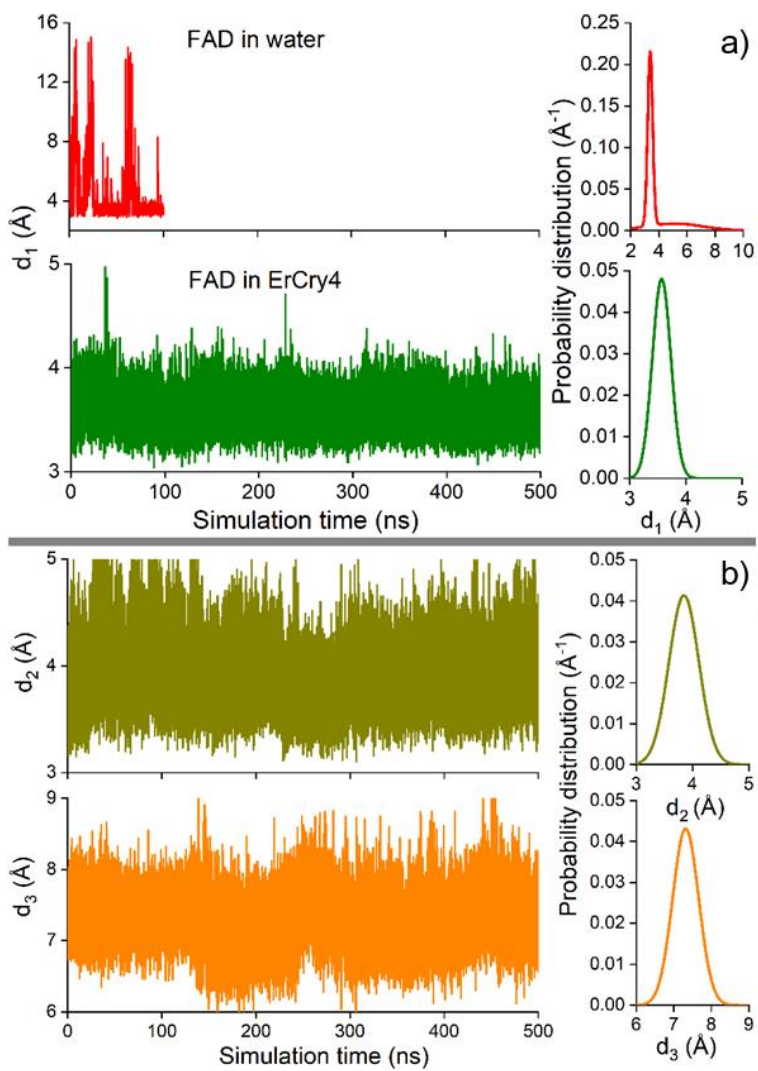

**Figure S9:** Edge-to-edge distances between the flavin (Fig. 5, blue) and adenine (Fig. 5, green) parts of the FAD cofactor as well as distances to nearby tryptophan residues which could play a role in alternative electron transfer routes in *Er*Cry4. a) Flavin-adenine edge-to-edge distance for solvated FAD in water and embedded in *Er*Cry4, respectively, with the corresponding probability density distributions. b) Edge-to-edge distances between adenine and the Trp290 residue ($d_2$) and flavin and the Trp350 residue ($d_3$) in *Er*Cry4, see Fig. 5. Time evolution of the two distances as well as the probability density distribution of the distances are shown.